# Nuclear matter properties, phenomenological theory of clustering at the nuclear surface, and symmetry energy

Q. N. Usmani[a*], Nooraihan Abdullah[a], K. Anwar[a] and Zaliman Sauli[b]

[a]Institute of Engineering Mathematics, University Malaysia Perlis, Malaysia

[b]School of Microelectronics Engineering, University Malaysia Perlis, Malaysia

## Abstract

We present a phenomenological theory of nuclei that incorporates clustering at the nuclear surface in a general form. The theory explains the recently extracted large symmetry energy by Natowitz *et al.* at low densities of nuclear matter and is fully consistent with the static properties of nuclei. In phenomenological way clusters of all sizes, shapes along with medium modifications are included. Symmetric nuclear matter properties are discussed in detail. Arguments are given that lead to an equation of state of nuclear matter consistent with clustering in the low density region. We also discuss properties of asymmetric nuclear matter. Because of clustering, an interesting interpretation of the equation of state of asymmetric nuclear matter emerges. As a framework, an extended version of Thomas Fermi theory is adopted for nuclei which also contain phenomenological pairing and Wigner contributions. This theory connects the nuclear matter equation of state, which incorporate clustering at low densities, with clustering in nuclei at the nuclear surface. Calculations are performed for various equation of state of nuclear matter. We consider measured binding energies of 2149 nuclei for $N, Z \geq 8$. The importance of quartic term in symmetry energy is demonstrated at and below the saturation density of nuclear matter. It is shown that it is largely related to the use of, *ab initio*, realistic equation of state of neutron matter, particularly the contribution arising from the three neutron interaction and somewhat to clustering. Reasons for these are discussed. Because of clustering the neutron skin thickness in nuclei is found to reduce significantly. Theory predicts new situations and regimes to be explored both theoretically and experimentally.

[*] qnusmani@hotmail.com

# 1. INTRODUCTION

There is mounting evidence [1], both theoretically and experimentally, that clustering ($^2H$, $^3H$, $^3He$, $\alpha$–particles and possibly heavier nuclei) occurs at the nuclear surface; a phenomenon which normally can not be described in Mean Field Theories. This clustering is evident from the studies of Natowitz *et al*. who have reported large values of symmetry energy at nuclear matter (NM) densities $\rho < 0.01$ fm$^{-3}$ at low temperatures [2,3]. The clustering arises because extra binding energies are gained because of cluster formation of various shapes and sizes in the equation of state (EOS), $E(\rho)$, of symmetric nuclear matter[1] (SNM) at subsaturation densities [4]. It finds explanation in Quantum Statistical (QS) [2, 4] approach which includes specific cluster correlations and then interpolates between the low density limit and the relativistic mean field (RMF) approaches near the saturation density. In QS approach only clusters with $A \leq 4$ has been included.

In the present study, we adopt a different approach. We introduce a thermodynamically consistent phenomenology in which we include the possibility of formation of clusters of all shapes and sizes along with medium modifications. We begin with the assertion that the binding energy of SNM per nucleon in the limit of zero density must approach $u_v$(≈16 MeV) at the saturation density $\rho_0$(≈0.16 fm$^{-3}$), where $u_v$ is the volume term in the Weizsäcker mass formula. The foregoing assertion is based upon the principle that nuclear matter in its ground state (T=0 MeV) at a given density will attain the lowest possible energy. Our assertion can be proven in the following way. It was demonstrated in Ref. [5] that an idealized $\alpha$-matter picture of SNM in the neighborhood of zero density gives energy per nucleon, $E/A \approx -7.3$ MeV, which is the energy of $\alpha$- particle per nucleon with Coulomb interaction switched off. The above conclusion is at variance with the mean field theories results where $E(\rho) \to 0$ as the density $\rho \to 0$, for example, in the Skyrme-Hartree-Fock calculations. But why $\alpha$-particles, why not heavier nuclei which have lower energies per nucleon compared to $\alpha$-particle? For example, an ideal $^{40}Ca$-matter will have lower *E/A* compared to an ideal $\alpha$-matter. This line of argument can be extended further, since heavier the nuclei lower are the *E/A* as the Coulomb interaction is switched off. We may thus consider SNM as an ideal gas of chunks of NM or very heavy nuclei in which we can ignore the surface effects. We consider the densities low enough so that interactions between clusters are negligibly small and the concept of ideal *cluster-matter* is valid. Clearly in such a situation, we obtain the exact result $E(\rho) \to -u_v$ as $\rho \to 0$. We therefore obtain the relation

$$E(\rho \to 0) \to E(\rho = \rho_0) = -u_v \qquad (1)$$

---
[1] By SNM we understand nuclear matter with equal numbers of neutrons and protons with Coulomb interaction turned off.

Relation (1) is counter intuitive and, to the best of our knowledge, has never been exploited in any nuclear physics calculations. We shall make extensive use of relation (1) in our formulation. At finite but low densities, the SNM is not an ideal cluster matter. The ideal cluster matter exists only in the limit of zero density. Thus nuclear matter at low densities will be a highly correlated system with complicated structure. Nonetheless, based on somewhat general considerations, we demonstrate in section II that EOS of SNM will have one maximum between 0 and $\rho_0$. This conclusion along with the identity (1) then includes the possibility of presence of clusters at low densities and leads to a compelling interpretation of the symmetry energy data of Natowitz *et al*. We postulate an EOS of SNM which contains the identity (1), a maximum between 0 and $\rho_0$, and explicitly other empirically found quantities, namely the compression modulus *K*, the saturation density and the volume term $u_v$. To determine the parameters of EOS we make use of the binding energy and charge *rms* radii data of nuclei; for which we use an approximate but well tested theory. One of the parameters of EOS is found sensitive to the symmetry energy data of Natowitz *et al.* [2]. This way we determine a semi-empirical EOS in the entire region between 0 and $\rho_0$ with the materialization of four new empirical quantities. These are the spinodal density $\rho_{sp}$ at which the homogeneous nuclear matter changes its character, $\rho_{min}$ the density at which symmetry energy shows a minimum, and the density $\rho_{max}$ at which the SNM attains a maximum value $E_{max}$.

Another novel feature of the present study is the use of neutron matter EOS obtained recently through *ab initio* calculations with realistic interactions [6-8]. It is demonstrated that these EOS give rise to a quartic isospin term in the symmetry energy for a consistent description of nuclei. Though it has been suggested [7] that these EOS can be incorporated, or more appropriately indicate, an adjustment of the coefficient of the isovector gradient term in Skyrme density functional. But, the isovector gradient term contributes mainly in the surface region whereas the quartic isospin term affects the entire region. It is shown that the quartic term is much more efficient; its inclusion considerably decreases or almost entirely eliminates the importance of the isovector gradient term.

Relation (1) should also hold for other quantum liquids, namely the two isotopes of helium $^3He$ and $^4He$ at zero temperature. They have some relevance to the present study, particularly $^4He$, since exact Green's Function Monte Carlo (GFMC) and Diffusion Monte Carlo (DMC) calculations exist for this system. Experimental data on these liquids also exist.

Incorporation of (1) into a theory is far from trivial. Clearly Hartree-Fock methods cannot take into account clustering at the nuclear surface since they do not satisfy (1); though mean field theories do contain some aspect of clustering but they are not related to surface properties. There are cluster models of nuclei and hypernuclei in terms of specific clusters, mostly $\alpha$-particle clusters, but these are confined to light nuclei [9]. On the other hand we have *ab initio* theories which give accurate or nearly exact results, but they are also confined to light nuclei [10]. In future, it may become possible to extend *ab initio* theories to heavy nuclei with realistic Hamiltonians, but it would be difficult to entangle cluster properties particularly its universal character, if any, at the nuclear surface. Recently, there has been a spurt of activity in developing microscopically based nuclear density functional theories [11-13] for medium and heavy nuclei with reference to realistic two- and three-nucleon or Skyrme interactions. The realistic interactions used are either the Argonne AV18 (two-body) [14] and Urbana IX (three-body) [15] or they are derived from chiral effective field theory. These studies have started with the motivation to search for a Universal Nuclear Energy Density Functional – a new model for nuclear theory [16], which in principle is supposed to have all the many body correlations. But none of these theories, because of constrains or choice of density functional, conform to identity (1); on the contrary they give $E(\rho)\rightarrow 0$ as $\rho\rightarrow 0$. They do not include the clustering aspect in nuclei to the full extent. Besides, on the basis of fits to energy alone it will be difficult to see the impact of (1) unless one probes those properties of nuclei which are sensitive to surface region such as the neutron skin thickness, the symmetry energy at low densities, and the one neutron separation energies for those pair of nuclei where separation energies are small.

To find a detailed structure of the EOS of SNM with GFMC/DMC methods at low densities is a formidable task and beyond the reach of present day techniques and computing power. We thus resort to phenomenology and approximate methods. We use an extended version of Thomas-Fermi model (ETF) which we believe is the only theory at present which can incorporate (1) easily and at the same time describe the properties of large number of nuclei. The ETF theory provides quick and reliable estimates of energies and other physical quantities of interest with a global insight. This implementation not only explains the experimental large values of symmetry energy [2] at low densities but also affect the binding energies of nuclei, one and two neutron separation, and $\beta$–decay energies in the right direction, though by a modest amount. But the influence of clustering on the neutron skin thickness and one neutron separation energies for those pairs of nuclei with separation energies less than 5 MeV are significant. The latter improvement is crucial for describing the nuclei near the drip line.

In spite of the use of an approximate theory, the conclusions drawn based upon the present approach will be of general validity and guide us to explore new circumstances and regimes. For example, because of cluster formation, the $\Lambda$-binding to SNM in the neighborhood of zero density must approach to its value at the saturation density which is around 30 MeV – an outcome of the conceptual requirement explained earlier. Implementation of this requirement shall constitute a fundamental departure from all the other earlier approaches [17-20] and requires a separate study. In addition to explaining the large values of symmetry energy [2] at low densities we find that it has a minimum (at $\rho_{min}$) between 0 and $\rho_0$, an interesting situation to be explored experimentally. Curiously, there is no experimental data in the neighborhood of $\rho_{min}$. The data exist at relatively higher and lower densities from $\rho_{min}$. Another important outcome is the necessity of quartic term. It is demonstrated that it largely originates from the use of realistic EOS of neutron matter, particularly the role played by three-nucleon interaction around $\rho \approx \rho_0$. It is also found somewhat connected with the clustering at the nuclear surface, a fact also observed in the microscopic QS approach [4].

Because of cluster formation, the neutron and proton densities at the nuclear surface would tend to equalize each other. This tendency to equalize results into reduced neutron skin thickness as compared to when clustering is absent. We do find considerable reduction in neutron skin thickness. It has bearings on the recently concluded parity violating electron scattering experiment on $^{208}Pb$ and in turn on the neutron star radii [21,22].

We perform large number of calculations pertaining to different aspects of EOS. These are carried out including 367 spherical nuclei. We then choose from these a few relevant ones for which calculations are performed for the measured binding energies of 2149 nuclei for $Z \geq 8$ and $A \geq 16$.

In section II, we discuss nuclear matter properties and postulate EOS of nuclear matter consistent with clustering. In section III, this EOS is included in an extended version of Thomas Fermi theory. We present our results in section IV with a detailed analysis. Section V gives conclusions and future outlook.

## II. NUCLEAR MATTER PROPERTIES

In density functional theories including the ETF theory, the equation of state of symmetric nuclear matter, $E(\rho)$, is an important quantity. It appears in these theories in an essential way. We therefore examine its properties in detail and find out ways through which the identity (1) can be included in the theory.

Empirically, we have information about three quantities relating to SNM. The first is the compression modulus $K$ whose value has been deciphered from Isoscalar Giant Monopole Resonances. We fix its values at $K=230$ MeV [23]. All our calculations are for this value of $K$. Our results are quite insensitive to the assumed value of $K$. We demonstrate this lack of dependence on $K$ by making one calculation for $K=260$ MeV. Other two empirical quantities, namely, $\rho_0 \approx 0.16$ fm$^{-3}$ and $u_v \approx 16$ MeV are sensitive to static properties of nuclei and symmetry energy. We vary these around their assumed values.

Next, the question arises, as how does $E(\rho)$ varies between zero and the saturation density. We assume that it is continuous and that its derivatives exist. We further assume that between 0 and $\rho_0$ it has one maximum. We advance arguments which are based on contradiction and inference to support this assumption. First, one may argue [24], that because of cluster formation at subsaturation densities, there are no maxima or minima, i.e., $E(\rho)$ is flat or independent of density between 0 and $\rho_0$. This conjecture implies that nuclear matter is infinitely degenerate; for each density between 0 and $\rho_0$ it has the same energy per particle, namely $-u_v$. There is no evidence for this assertion. On the contrary, it is contradicted by all the SNM calculations with realistic interactions [25-27]. In particular we refer to the Auxiliary Field Diffusion Monte Carlo Calculations (AFDMC) of Gandolfi *et al*. [27]. AFDMC generates correlations needed to form clusters. From the point of view of many-body problem, these are accurate calculations and contain important *A*-body correlations generated by the interaction. A clear minima in $E(\rho)$ is evident (Fig. 1 of Ref. [27]) as a function of density at a saturation density with no flattening in the subsaturation region. Rather clean evidence against the conjecture comes from liquid $^4He$. Unlike nuclear matter, a hypothetical system, liquid $^4He$ is a real self bound saturating system. Exact GFMC and DMC calculations with accurate quantum mechanical Hamiltonian give results indistinguishable from experiment [28, 29]. With the same Hamiltonian liquid $^4He$ droplets, in various respects analogous to atomic nuclei, have been studied with the exact GFMC/DMC methods for $3 \leq N \leq 112$ where *N* is the number of $^4He$ atoms [30, 31]. These droplets are bound and their energies can be well reproduced with a liquid–drop formula containing volume, surface and curvature terms. Therefore, according to the conjecture, liquid $^4He$ should breakdown into clusters of $^4He$ droplets below the saturation density with energy per particle of the "broken liquid" equal to its value at the saturation density. But this breaking is not found through calculations.

Calculations show clear minima in energy at the saturation density when plotted as a function of density. In Fig 1, we summarize the results of various GFMC and DMC calculations of the energy par particle as a function of density. The down and up triangles are the GFMC calculations of Kalos *et al.* [28] with the Lennard-Jones [32] and the HFDHE2 potentials [33] respectively. The dashed and long dashed lines are drawn to aid the eye. The solid circles are the experimental values from Refs [34]. The open circles are the experimental values obtained after making correction to zero temperature, Ref. [35]. The solid line represents the fit to DMC calculations [29] with the HFD-B(HE) potential [36]. The curve accurately represents the results from the spinodal density (0.264 $\sigma^{-3}$) right up to the highest density of 0.650 $\sigma^{-3}$ ($\sigma$=2.556 Å). A flat $E(\rho)$ is ruled out for liquid $^4He$ and thus in turn for nuclear matter. Next, assume on the contrary, that not one but two maxima in $E(\rho)$ exist in the subsaturation region. The two maxima will surround a minimum at an intermediate density $\rho_{in}$. This $\rho_{in}$ will then be an additional saturation or equilibrium density where nuclear matter will be stable or metastable, though it may have higher *E/A*. It has been shown in the past that if one assumes two saturation densities, one at the normal nuclear matter saturation and the other at a higher density, separated with a saturation barrier, one will find collapsed nuclei corresponding to saturation at higher densities [37]. In our case, the additional saturation density $\rho_{in}$ (< $\rho_0$) separated with a barrier from $\rho_0$ will then lead to formation of extended nuclei in metastable state with larger rms radii, since $\rho_{in}$ is at a lower density. We have no experimental evidence for these extended nuclei. We therefore conclude that $\rho_{in}$ does not exist. We believe it is a safe conclusion. Even if a $\rho_{in}$ exists such that the difference between the *E/A* values of the minimum and one of the surrounding maximum is small, it will escape experimental detection of metastable states in nuclei mentioned above. But then its effect on the known nuclear properties is expected to be small; we ignore it. If we combine the foregoing remarks and observations with the identity (1), it follows that a single maximum in $E(\rho)$ between 0 and $\rho_0$ is a reasonable assumption and contains the essential physics that we wish to study.

Thus an EOS which satisfies (1) with one maximum in the subsaturation region will contain all possible clusters in their most general form. The contribution of individual clusters from deuteron to heavy nuclei will determine the shape of EOS as a function of density. In determining the shape of EOS we shall be guided by the symmetry energy data at low densities [2] and the static properties of nuclei [38, 39].

To sum up, we arrive at the following picture of SNM. At the saturation density the SNM is stable with $-u_v$ MeV of energy per nucleon. At lower densities, in the neighborhood of $\rho_0$, nuclear matter will be in a stretched state with higher energies in compliance with Ref. [25-27] till we reach the spinodal density at which the compression modulus becomes zero. Between the spinodal and saturation density, one may expect that nuclear matter can exist in metastable states with long lifetime as it has been observed for real liquids, for example the two isotopes of liquid helium and classical liquids such as water [40]. Below the spinodal density nuclear matter becomes thermodynamically unstable. Here, one considers, as suggested by various calculations [4, 41, and 42], formation of cavities, nuclei of exotic shapes and clusters and other possible phases of nuclear matter. We are not concerned here in the details of the structure of this region. Our interest lies in the behavior of $E(\rho)$ as a function of $\rho$. In the absence of any unswerving knowledge below this density we shall assume that energy still rises up to some density, neglecting possible small fluctuations, as the density is lowered. Further lowering will eventually bring us to some density where the energy per nucleon will be a maximum. Existence of this maximum, as we shall see in Sect. IV, leads to a minimum in the symmetry energy which should be verifiable in future experiments. Onset of clustering in the nuclear matter may begin right below the spinodal density and continue till the density at which energy is maximum[2]. With lowering densities below $E_{max}$ (at $\rho_{max}$) one can now envision merging of clusters instead of its creation since pressure is positive. In this region, SNM gives away its energy by performing external work. This process can be continued with the formation of larger clusters and thus lower energies per nucleon, till we reach the zero density. In that limit $E(\rho\rightarrow 0)$ again becomes $-u_v$ MeV and pressure zero in accordance with relation (1). Our EOS for SNM adheres to this picture. As mentioned earlier, in this region of densities below the spinodal density, nuclear matter is thermodynamically unstable. But, in nuclei, low densities of nuclear matter become stable because of surface effects and Coulomb interaction plays significant role.

We shall use these concepts in formulating our phenomenological EOS of SNM at T=0 MeV. Our approach in this study is macroscopic, which will hide and/or not require microscopic details such as in QS and other approaches [2-4]. Thus effect of all the specific clusters from deuteron to heavy nuclei, their condensates along with medium modifications are assumed contained in our EOS of SNM which, as we shall see, greatly impacts the symmetry energy at low densities. In actual nuclei, the effect of the clustering will be modified because of *N-Z* asymmetry, Coulomb forces which would prefer smaller cluster sizes, and finite number of nucleons. These effects are included in our formulation.

To be consistent with [2] and ignoring small effects because of charge symmetry breaking, the symmetry energy, $E_{sym}(\rho)$, is defined as the difference between $E(\rho)$ of pure neutron matter (PNM) and the SNM:

---

[2] In nuclei with Coulomb forces or in nuclear matter with Coulomb interaction turned on (maintaining the charge neutrality with electron fraction) the clustering may begin right below the saturation density [41]. But our discussion at the moment is without the Coulomb forces applicable only for the hypothetical nuclear matter; they are added finally to correctly depict the real picture in nuclei.

$$E_{sym}(\rho) = E(\rho, \delta = 1) - E(\rho, \delta = 0) \qquad (2)$$

where $\delta = (\rho_n - \rho_p)/\rho$ is the isospin asymmetry; $\delta=1$ (PNM), $\delta=0$ (SNM). $\rho_n$ and $\rho_p$ are the densities of neutrons and protons respectively and $\rho = \rho_n + \rho_p$ is the total baryon density. Since, neutron matter is a gas of neutrons at all densities with zero binding energy per nucleon at zero density, we get the exact result $E_{sym}(\rho \to 0) \to u_v$.

We shall denote by $E(\rho, \delta)$ the energy of NM per nucleon with isospin asymmetry $\delta$. Expanding $E(\rho,\delta)$ by Taylor series, keeping terms up to the fourth power of $\delta$ and ignoring the small charge symmetry breaking component, we obtain

$$E(\rho, \delta) = E(\rho,0) + S(\rho)\delta^2 + Q(\rho)\delta^4 \qquad (3)$$

where the factorials have been absorbed in $S(\rho)$ and $Q(\rho)$ and $E(\rho) \equiv E(\rho,0)$. At present, we have no information about the density dependence of $Q(\rho)$. We make the simplifying assumption that $Q(\rho)$ is proportional to $S(\rho)$. This assumption leads to satisfactory fit to the static properties of nuclei. We therefore replace $S(\rho) = (1-q) E_{sym}(\rho)$ and $Q(\rho) = q E_{sym}(\rho)$. The parameter $q$ will be density dependent if $S(\rho)$ and $Q(\rho)$ have different density dependence. Since, we have assumed otherwise $q$ will be independent of density. The parameter $q$ determines the relative importance of the two terms and is found to play an important role in giving good fits to the static properties of nuclei. With these replacements, Eq. (3) assumes the form

$$E(\rho, \delta) = E(\rho) + E_{sym}(\rho)\left((1-q)\delta^2 + q\delta^4\right) \qquad (4)$$

Substituting for $E_{sym}(\rho)$ from (2) into (4) then leads to

$$E(\rho, \delta) = E(\rho, \delta = 0) + \left(E(\rho, \delta = 1) - E(\rho, \delta = 0)\right)\left((1-q)\delta^2 + q\delta^4\right) \qquad (5)$$

It is seen that the dependence on the isospin variable $\delta$ occurs only in the expression $(1-q)\delta^2 + q\delta^4$ on the *rhs* of (4) and (5). For $\delta = 1$, the *rhs* of (5) then trivially gives the EOS of pure neutron matter and for $\delta = 0$, it gives the EOS of SNM.

In the past, mainly the Skyrme density functionals have been employed in the Extended Thomas Fermi theory [43-46]. They do not satisfy the identity (1). We therefore use quite general density functional for $E(\rho)$ containing the constrain (1) and a maximum between 0 and $\rho_0$ as explained earlier. Fulfilling this requirement is consistent with DFT theories, where effective interactions are secondary to the theory; it is the density functional that defines or should define the force [16]. Following Ref [17] we write

$$E(\rho_\geq) = -u_v + \frac{K}{18}\left(\frac{\rho - \rho_0}{\rho_0}\right)^2 + M\left(\frac{\rho - \rho_0}{\rho_0}\right)^3, \qquad \rho \geq \rho_x \quad (6a)$$

where, $\rho_x$ is a parameter between 0 and the saturation density $\rho_0$. For $\rho \leq \rho_x$ we consider two options

$$E(\rho_\leq) = -u_v + A\rho + B\rho^2 + C\rho^3 + D\rho^4 + \frac{3\hbar^2(3\pi^2/2)^{2/3}}{5m_N}\rho^{2/3} \qquad \rho \leq \rho_x \quad (6b)$$

$$E(\rho_\leq) = -u_v + A\rho + B\rho^2 + C\rho^3 + D\rho^4. \qquad \rho \leq \rho_x \quad (6c)$$

Notice in (6b) and (6c), when the density approaches zero, the binding energy per nucleon becomes $u_v$ in agreement with (1). In (6c) there is no Thomas-Fermi kinetic energy term (the term proportional to $\rho^{2/3}$). One may take the point of view that this kinetic energy is indirectly contained in (6c) through terms proportional $\rho^n$ for n =1 to 4. In fact, through fitting of the data, it turns out that (6b) is only marginally preferred over (6c). Our $E(\rho)$ follows the general pattern as a function of density as explained earlier in detail. The constant terms *A*, *B*, *C* and *D* are determined by equalizing $E(\rho_\geq)$ and $E(\rho_\leq)$ and their first three derivatives at $\rho = \rho_x$. The equation of state (6) contains mainly two parameters $M$ and $\rho_x$. The other parameters are consistent with the generally accepted values, $u_v \approx 16$ MeV, the compression modulus $K \approx 230$ MeV and $\rho_0 \approx 0.16$ fm$^{-3}$. As stated earlier, we do vary $u_v$ and $\rho_0$ around these values, $K$ is fixed at 230 MeV.

We shall also consider the case when there is no clustering, achieved by putting $u_v=0$ in (6b). Thus (6b) modifies to

$$E(\rho_\leq) = A\rho + B\rho^2 + C\rho^3 + D\rho^4 + \frac{3\hbar^2 (3\pi^2/2)^{2/3}}{5m_N}\rho^{2/3} \ . \quad \text{For } \rho \leq \rho_x \quad (6d)$$

In this context, we have not considered here modification to (6c) since it does not lead to anything new as observed in the previous paragraph. (6d) can be interpreted as a low density expansion. The term proportional to $\rho$ may be considered as arising out of the two-body interaction. The rest of the terms may arise because of three and many-body interactions and correlations. Such an interpretation is not possible for (6b) and (6c) because of the term $u_v$ which embodies strong correlations and clustering at low densities. (6d) will be in line or close to Skyrme-Hartee-Fock or other mean field theories functionals. Densities greater than $\rho_0$ are not relevant in the present study. They are not accessible or reached in nuclei except when shell corrections are incorporated.

**III. AN EXTENDED THOMAS FERMI MODEL**

We adopt an extended version of Thomas-Fermi (ETF) method which is based on the density functional approach [17, 43-46]. This technique has been extensively used in atomic, metallic clusters and nuclear physics (see M. Brack *et al*. Ref. [43]) and accurately produces the average part of the energy. Thus the quantal shell effects are smoothed out as in the liquid drop model or Strutinsky's [47] calculations. Here we use a simplified version of the extended Thomas –Fermi theory where we ignore the spin-orbit densities. It is found sufficient for the present exploratory study. Its inclusion is left for future and is expected to further improve the results. But its effects are somewhat approximately included through shell correction. We assume spherical symmetry. We write the energy of a nucleus, $\mathcal{E}[\rho_n, \rho_p]$, as a functional of the neutron and proton densities:

$$\mathcal{E}[\rho_n,\rho_p] = \int \left[ E(\rho,\delta)\rho + \frac{\hbar^2}{2m}\left(\tau_2(\rho_n,\rho_p) + \tau_4(\rho_n,\rho_p)\right) + a_\rho(\nabla\rho)^2 - a_{np}(\nabla\rho_n - \nabla\rho_p)^2 \right] d\vec{r}$$
$$+ \frac{1}{2}e^2 \int \frac{\rho_p(\vec{r})\rho_p(\vec{r}\,')}{|\vec{r}-\vec{r}\,'|} d\vec{r}\,' d\vec{r} - \frac{3}{4}\left(\frac{3}{\pi}\right)^{1/3} e^2 \int \rho_p^{4/3}(\vec{r}) d\vec{r}$$
$$+ Shell + a_{pair} A^{-1/3} \Delta_{np} + E_W$$

If we use relation (4) for $E(\rho,\delta)$ in the above expression, we get

$$\mathcal{E}[\rho_n,\rho_p] = \int \left[ E(\rho)\rho + \frac{\hbar^2}{2m}\left(\tau_2(\rho_n,\rho_p) + \tau_4(\rho_n,\rho_p)\right) + a_\rho(\nabla\rho)^2 - a_{np}(\nabla\rho_n - \nabla\rho_p)^2 \right] d\vec{r}$$
$$+ \int \left[(1-q)E_{sym}(\rho)\delta^2 + qE_{sym}(\rho)\delta^4\right] \rho d\vec{r} \tag{7}$$
$$+ \frac{1}{2}e^2 \int \frac{\rho_p(\vec{r})\rho_p(\vec{r}\,')}{|\vec{r}-\vec{r}\,'|} d\vec{r}\,' d\vec{r} - \frac{3}{4}\left(\frac{3}{\pi}\right)^{1/3} e^2 \int \rho_p^{4/3}(\vec{r}) d\vec{r}$$
$$+ Shell + a_{pair} A^{-1/3} \Delta_{np} + E_W$$

where $\tau_2$ and $\tau_4$ are the kinetic energy densities

$$\tau_2(\rho_n,\rho_p) = \sum_{k=n,p} \left( \frac{1}{36} \frac{(\nabla\rho_k)^2}{\rho_k} + \frac{1}{3}\Delta\rho_k \right) \tag{8}$$

$$\tau_4(\rho_n,\rho_p) = \frac{1}{6480}(3\pi^2)^{-2/3} \sum_{k=n,p} \rho_k^{1/3} \left( 8\left(\frac{\nabla\rho_k}{\rho_k}\right)^4 - 27\left(\frac{\nabla\rho_k}{\rho_k}\right)^2 \frac{\Delta\rho_k}{\rho_k} + 24\left(\frac{\Delta\rho_k}{\rho_k}\right)^2 \right) \tag{9}$$

The summation $k$ in (8) and (9) is over the neutron and proton densities. The $\hbar^2/2m$ factor that multiplies $\tau_2$ and $\tau_4$ in (7) contains the bare nucleon mass $m$. The integral in the first line of (7) can be interpreted [43] as the volume and surface terms, the second integral is the contribution due to symmetry energy and the last two integrals are respectively the direct and exchange Coulomb energy in the Slater approximation. The term $a_\rho(\nabla\rho)^2$ and $a_{np}(\nabla\rho_n - \nabla\rho_p)^2$ are because of the finite range of the interaction and control the surface properties. They have been introduced taking a cue from Skyrme density functional [42]. The parameter $a_{np}$ multiplying $(\nabla\rho_n - \nabla\rho_p)^2$ has been the subject of an interesting recent *ab initio* study of drops of neutron matter trapped in an external field [6]. We find that this term is rendered ineffective by the quartic isospin term in the symmetry energy. We shall return to this discussion in the next section. In the last line we have the quantal shell contribution which we extract from Ref [48]. The shell contributions also include deformation energies. The last two terms are the pairing energy and Wigner ($E_W$) contributions, respectively.

For $\Delta_{np}$, following [49, 50], we use

$$\Delta_{np} = \begin{cases} 2-|I| : \text{N and Z even} \\ |I| : \text{N and Z odd} \\ 1-|I| : \text{N even Z odd and N} > \text{Z} \\ 1-|I| : \text{N odd Z even and N} < \text{Z} \\ 1 : \text{N even, Z odd and N} < \text{Z} \\ 1 : \text{N odd, Z even and N} > \text{Z} \end{cases} \quad (10)$$

with $|I| = |N-Z|/A$.

We use a phenomenological Wigner term

$$E_W = V_W \exp\left\{-\lambda\left(\frac{N-Z}{A}\right)^2\right\} + W_W |N-Z| \exp\left\{-\left(\frac{A}{A_0}\right)^2\right\} \quad (11)$$

This form has been proposed by Goriely *et al.* [51] in their Skyrme-Hartree-Fock-Bogoliubov microscopic-macroscopic mass formula.

The shell, pairing and Wigner terms do not play significant roles, as far as the present study is concerned, but they improve the results, i.e. fits to energies and rms radii, quantitatively, and make us compare our results with other microscopic-macroscopic theories. Their inclusion is on the same line as in the liquid drop model.

For variational neutron (proton) densities, we employ three parameter modified Fermi distribution for each species [43,52]:

$$\rho_{n(p)}(r) = N_{n(p)} / \left(1 + \exp((r - R_{n(p)})/t_{n(p)})\right)^{\gamma_{n(p)}} \quad (12)$$

with $R_{n(p)}$, $t_{n(p)}$ and $\gamma_{n(p)}$ as variational parameters for neutrons(protons) and $N_{n(p)}$ is the normalization constant ensuring the correct neutron (proton) numbers. For each nucleus we vary the six parameters to minimize the energy.

A positive feature of the present approach is that the microscopically calculated homogeneous EOS of neutron matter can be directly employed. We use the recently calculated values [7, 8]. This has been obtained by employing an accurate fixed phase AFDMC technique with 66 neutrons enclosed in a periodic box with Argonne AV8' [14] and Urbana three-nucleon UIX [15] interactions. There is little difference between the results of neutron matter for AV8' and AV18' in the low density region. But, the results may differ if the more realistic Illinois (IL) three-body interaction [53] is used required for producing the ground and excited state energies of p-shell nuclei in the GFMC calculations [10]. In Fig. 2 (left panel) we plot the results of Ref. [7], represented by filled circles for AV8'+ UIX. The solid line is the fit obtained by

$$E(\rho) = \sum_{i=1}^{3} y_i \rho^i \bigg/ \left(1 + \sum_{i=1}^{4} z_i \rho^i\right) \quad (13)$$

where the parameter values are $y_1 = 0.331 \times 10^4$ MeV fm$^3$, $y_2 = 0.632 \times 10^7$ MeV fm$^6$, $y_3 = 0.259 \times 10^9$ MeV fm$^9$, $z_1 = 0.948 \times 10^4$ fm$^3$, $z_2 = 0.192 \times 10^7$ fm$^6$, $z_3 = 0.772 \times 10^7$ fm$^9$, $z_4 = -0.323 \times 10^8$ fm$^{12}$. The open circles represent the results with AV8' alone and can be obtained by multiplying the solid curve with a fudge factor $exp(-2.615(\rho-0.05))$ for $\rho > 0.05$ fm$^{-3}$. We use these fits in our calculations of $E_{sym}(\rho)$. Gandolfi et al. [7] have given a different parameterized form for $E(\rho)$ of neutron matter

$$E(\rho) = a\left(\frac{\rho}{\rho_0}\right)^{\alpha} + b\left(\frac{\rho}{\rho_0}\right)^{\beta} \qquad (14)$$

We also use this form in one particular case, but (13) gives a better parameterization for $\rho \leq 0.04$ fm$^{-3}$.

We define the root mean square deviations for energies and radii

$$\sigma(E) = \sqrt{\sum_{i=1}^{N}\left(E_{theo}^i - E_{\exp}^i\right)^2 / N} \qquad (15a)$$

with a similar definition for $\sigma(R)$,

$$\sigma(R) = \sqrt{\sum_{i=1}^{N'}\left(R_{theo}^i - R_{\exp}^i\right)^2 / N'} \qquad (15b)$$

where $N$ and $N'$ are the numbers of nuclei included. The *rms* deviations $\sigma(S1), \sigma(S2),$ and $\sigma(Q_\beta)$ for the one and two neutron separation and $\beta$–decay energies respectively are defined analogously.

## IV RESULTS AND DISCUSSION

Organization of this section is as follows. First we describe the results with a smaller set of 367 spherical nuclei. We have made calculations for a number of situations, 17 in number, corresponding to different EOS (6b-d), neutron matter with and without three-body interaction and different values of $\rho_x$ since symmetry energy is sensitive to this parameter. Out of these we narrow down a few relevant ones, plus a few additional ones, for which calculations are carried out for 2149 nuclei [38] for $N, Z \geq 8$. Next we present results for the symmetry energy between 0 and $\rho_0$, EOS of asymmetric nuclear matter, and neutron skin thickness and other relevant quantities. Discussion follows the results as they are described. Finally, we present an overall critique of the results.

**A. Calculations for Nuclei:** To begin with, to make computations relatively less extensive, we confine to 367 spherical nuclei [54] from $^{38}Ca$ to $^{220}Th$. They include the chains $^{38-52}Ca$, $^{42-54}Ti$, $^{100-134}Sn$ and $^{178-214}Pb$. These are the same set of nuclei considered by Bhagwat et al. [54]. The electronic binding energy $1.433\times10^{-5}Z^{2.39}$ MeV has been subtracted from the binding energies of Ref [38]. For the charge *rms* radii we have considered all the 149 nuclei of the 799 nuclei of Ref [39] common to the set of 367 nuclei. The point proton *rms* radii, $r_p$, are obtained through the relation $r_p = \sqrt{r_c^2 - 0.64}$, where $r_c$ is the charge *rms* radius.

We have a total of twelve parameters (which eventually is reduced to eleven), namely, $u_v$, $\rho_0$, $M$ and $\rho_x$ in (6), $a_\rho$ and $a_{np}$ which controls the surface, $q$ and $a_{pair}$ in (7), and the four parameters $V_W$, $\lambda$, $W_W$ and $A_0$ in the Wigner term (11). Calculated energies are obtained variationally by varying the density through the six variational parameters of (12). First we put $q = 0$, i.e. no isospin quartic term, and minimize $\sigma(E)$, (15a), for fixed values of $\rho_x$, varying the other parameters. Minimization is achieved through an automated search procedure. In Table I, we give results for $\rho_x = 0.06$ fm$^{-3}$. Other values of $\rho_x$ in the range $0.04 \leq \rho_x \leq 0.09$ give similar results, but, results for $\rho_x = 0.06$ fm$^{-3}$ give a better account of the symmetry energy – hence the choice. The neutron matter EOS employed is for the Hamiltonian AV8'+UIX. The EOS of SNM contains the clustering through (6b) and (6c). The parameter $a_{np}$ has significant effect on the *rms* deviation $\sigma(E)$. Also noteworthy are the large values of Wigner parameters compared to Ref. [51]; they also help lowering $\sigma_E$. Though the *rms* deviation $\sigma(R)$ is reasonable, but from the point of view of microscopic-macroscopic theories $\sigma(E)$ is quite large. Its average value is $\approx$ 1.4 MeV. Current values for this quantity lie in the range 0.5-0.7 [44, 50, 51,54-56].

In Table II we give the results with the quartic isospin term, i.e. $q$ is varied along with other parameters. The parameter $a_{np}$ was found very close to zero. Therefore we put it zero since it makes insignificant changes in $\sigma(E)$ when the quartic isospin term is included. All subsequent calculations (results of Table III to VI) bear this out without exception. We obtain a dramatic reduction in $\sigma(E)$, by more than a factor of 2 compared to the average value of 1.4 MeV from Table I. The improved values of $\sigma(E)$ is ≤ 0.6 MeV in line with the current microscopic-macroscopic theories. Wigner parameters also become reasonable. This amply justifies the inclusion of the quartic isospin term. In this and other subsequent tables, III to VI, we give the coefficients $(1-q) E_{sym}(\rho)$ and $qE_{sym}(\rho)$ which multiplies the $\delta^2$ and $\delta^4$ isospin terms respectively at the saturation density. Its relevance shall be discussed later on. We also included term proportional to $\delta^6$ but no significant change in $\sigma(E)$ was found. We shall discuss the reason for this fact at a later stage. To test the sensitivity of $\sigma_E$ with respect to $q$, in Fig 3 we plot $\sigma_E$ as a function of $q$. This curve was obtained by taking specific values of $q$ then minimizing $\sigma(E)$ with respect to other parameters. For neutron matter AV8'+UIX was used. SNM is with (6b). The figure demonstrates that $\sigma(E)$ is sensitive to $q$.

The appearance of quartic isospin term in fits to binding energies is something new, whereas in earlier microscopic-macroscopic theories this term is not needed. Why is that so? We shall elaborate on this fact after presenting results for other choices of EOS of neutron matter and situation pertaining to no-clustering (6d) in the EOS of SNM.

In Table III, results for other values of $\rho_x$ are given. The asterisk * denotes a modified 3$N$ interaction [8] motivated by the UIX [15], the Illinois models [53] and the symmetry energy $E_{sym}(\rho_0)$. The results of *ab initio* calculations of neutron matter EOS with one such particular model, named as $V_{2\pi}^{PW} + V_{\mu=300}^{R}$ [8], is obtained by putting $a = 12.8$ MeV, $\alpha = 0.488$, $b = 3.19$ MeV and $\beta = 2.20$ in (14).

We observe that $\sigma(E)$ is weakly dependent on $\rho_x$ in the range 0.04 to 0.07 fm$^{-3}$. Also the $q$ values for AV8'+UIX change very little. But for $AV8'+V_{2\pi}^{PW} + V_{\mu=300}^{R}$, it reduces by a factor of 2. The quantity $(1-q)E_{sym}(\rho_0)$ which multiplies $\delta^2$ in (4) and (7) changes by small amount, the change comes in $qE_{sym}(\rho_0)$ which multiplies the quartic isospin ($\delta^4$) term. A similar situation is seen in results with AV8' alone, Table IV. Here too $q$ value is smaller by a factor of 2 compared to values for AV8'+UIX (Table II and III) with similar trends for the quadratic and quartic isospin terms in the symmetry energy.

Table V gives results without clustering, i.e., with (6d). It is seen from table II-V that decrease in $\sigma(E)$ because of clustering is 8.5%, 7.8% and 6% for AV8'+UIX, AV8' and AV8'+$V_{2\pi}^{PW}$+$V_{\mu=300}^{R}$ respectively. The value of $q$ reduces when we switch to no-clustering situation, Table V, by ≈9% and 25% for AV8'+UIX and AV8', respectively. Considering the sensitivity of $\sigma_E$ on $q$, Fig. 3, this reduction is significant. Also we have given in Table V, result for AV8' (no-clustering) with $q = 0$ but include $a_{np}$. This demonstrates that quartic isospin term can be ignored when $q$ values are small, the isovector gradient term brings σ(E) close, but not lower, to the second line of results where $q = 0.062$. We still need non-zero values of $q$ for σ(E) to be same as in the second line of results. Here, the situation is not as bad as in the case of Table I, where we had clustering and a stiff EOS of neutron matter because of three-nucleon interaction. Thus if we have no clustering and softer EOS of neutron matter an isovector gradient term may suffice with no quartic isospin term as is the case with Skyrme density functional theories [6].

The Brussels-Montreal group [56] has reported that when they tune the Skyrme density functional parameters to mimic the APR [26] EOS of neutron matter, obtained with AV18+UIX, their quality of fits to masses deteriorates. This conclusion is in agreement with the results of Table I. UIX induces large quartic isospin term which cannot be emulated by the isovector gradient term as is evident from Tables I and II. The Brussels-Montreal group uses an older EOS of neutron matter by Friedman-Pandharipande (FP) [57], represented by the down triangle in the left panel of Fig. 1. It is less steep or repulsive compared to other EOS in the figure. Though we have not made calculations with this EOS, we believe that to a good approximation it will not require a quartic isospin term when clustering is not included as explained at the end of the previous paragraph. The EOS of neutron matter for AV8' and FP are close, Fig. 1 (left panel), though still AV8' is steeper than FP. The conventional Skyrme density functionals will be adequate in this situation. This answers the question raised earlier that why the quartic isospin term is not needed in earlier microscopic-macroscopic theories.

We now present calculations for 2149 nuclei [38]. These nuclei have regions of large deformation and also to some extent large |N–Z|/A values. Indeed, the extent of predicted deformation depends on the features of the density functional. In our case, the density functional (7) is spherical. Any contribution which arises from deformation is contained in the shell-correction contribution which we have taken from Ref. [48]. But, it would still be useful to perform these calculations as to see how the present approach with its limitation of self-consistency confronts the experimental data. We present in Table VI results for a limited set of EOS of symmetric and pure neutron matter. All the calculations are for $\rho_x = 0.06$ fm$^{-3}$. We used (6b) for the EOS of SNM; since as seen from Tables I to V that (6b) has a very slight edge over (6c). It is clear from Table VI that increase in $\sigma(E)$ is not very large compared to results for 367 spherical nuclei. On the average it increases by 0.08 MeV, still within reasonable limits from the viewpoint of microscopic-macroscopic theories. But $\sigma(R)$ increases by a factor of 2. We do not have any clear explanation for this discrepancy. We may attribute this increase to the absence of self-consistency and particularly to deformation in our formulation. We also give the

*rms* deviations σ(*S*1), σ(*S*2) and σ(*Q*$_β$) for one (*S*1) and two (*S*2) neutrons separation, and *β*-decay energies, respectively. Correct prediction of one and two neutron separation energies is important from the point of view of neutron drip line. One neutron separation energies govern the asymptotic density of neutrons [58] whereas the two neutron separation energies reveal the shell structure in an isotopic chain [54]. It is seen from Table VI that these quantities namely, σ(*S*1), σ(*S*2) and σ(*Q*$_β$), are quite reasonable. They are somewhat larger for the no-clustering case (line 2) implying that clustering does affect the fits though quite modestly; it was more so in case of 367 spherical nuclei where the difference is somewhat significant. Line 5 gives the results with Wigner term excluded. The value of σ(*E*) increases by 0.12 MeV. This demonstrates the importance of Wigner term. Our preferred results are for AV8' + $V_{2\pi}^{PW}$ + $V_{\mu=300}^{R}$ (line 3) and AV8' (line 4) since the *rms* deviations for these are small compared to other results. Last line of the Table VI give results for calculation with *K* = 260 MeV which demonstrates that dependence on *K* is weak. It is largely compensated by a change in the parameter *M*.

In Table VII, we compare our best fits with various other approaches. Column 4 gives the result from Ref. [55] of Möller *et al*. in the finite range droplet model. Column 5 and 6 are the results of Brussels-Montreal group in the Skyrme-Hartree-Fock-Bogoliubov microscopic-macroscopic approach [56, 59]. Column 7 is the result of Wang *et al*. [50] where they use an isospin dependent symmetry energy coefficient within the liquid drop model. Column 8 gives the result of Liu *et al*. [60], a significant improvement over the results of Ref. [50] obtained by the mirror nuclei constraint in the shell correction [61] and with some additional empirical residual shell corrections (for details see Ref. [60]). The last column is the result of Bhagwat *et al*. [54] in the liquid drop model with shell correction implemented through Wigner-Kirkwood expansion. The first line compares $\sigma(E)$, which is satisfactory. The second line compares $\sigma(R)$, about which we have commented earlier. The third line gives the *rms* deviation, $\sigma(S1)$, for one neutron separation energies (1988 measured values). Except that of Wang *et al*. [50] and Liu *et al*. [60] our values are lower. Line four gives result for two-neutron separation energies for the set of 1937 nuclei. Unfortunately we do not have numbers to compare from other studies, but its low value of 0.505 MeV does demonstrate that the shell structure as revealed in the isotopic chains is satisfactory. In line five, we give the result for the subset of 186 neutron rich nuclei for which one neutron separation energies are < 5 MeV. As is evident the *rms* deviation $\sigma(S_{nr})$ is significantly lower than other approaches. Later, when we present results for asymmetric nuclear matter, we shall discuss the significance of this result. In line six the *rms* deviation for 1868 measured *β*-decay energies is given. Here too our value is significantly lower than Brussels-Montreal group. Last line gives the number of nuclei considered in various approaches. It is noted that the deviations σ(*E*) of Wang *et al*. [50] and Liu *et al*. [60] are significantly lower than our values. But this comparison should be made in a proper perspective. Firstly, our microscopic part including the deformation is approximate and lack self consistency which can play an important role in bringing down σ(*E*). Secondly, references [50] and [60, 61] contain additional phenomenological terms in the shell-correction and Wigner term which significantly affect σ(*E*). These terms are not included in our formulation. Our aim here is to explain the symmetry energy data at low densities crucially linked with the clustering in nuclear matter, a phenomenon which can not be described in the framework of references [50, 60, and 61] and other mean field theories. In addition, in our formulation connection to realistic *NN* and *NNN* interactions is strong and direct where as in liquid drop models such as in references [50, 60, 61] this association is lacking. We primarily seek a reasonable description of the static properties of nuclei which we find satisfactory. We emphasize more on the physics related to symmetry energy and clustering rather than obtaining a precise fit to binding energies and other properties. We believe that improvements in the *rms* deviations $\sigma$ shall be achieved in future when we further improve our formulation including some of the effects considered in references [50, 60, and 61]. In Fig. 4, we plot the differences between the calculated (*cal*) and experimental (*exp*) energies (2149 nuclei, left panel) and the proton *rms* radii (773 nuclei, right panel). These values are plotted for the results given in line 4 of the Table VI. For energies, lighter nuclei show larger scatter compared to heavier nuclei. For *rms* radii this trend is much less pronounced.

It is seen from Tables II-VI, the quantity $(1-q)E_{sym}(\rho_0)$ varies between $\approx 28$ to 30 MeV whereas $qE_{sym}(\rho_0)$, which multiplies $\delta^4$, varies considerably; it is in the range of $\approx 2$ to 6 MeV. The quantity $(1-q)E_{sym}(\rho_0)$ which multiplies the $\delta^2$ term in isospin has often been denoted in the literature by the symbol $J$ [55, 56]. Its value is in agreement with other studies [50, 55, and 56]. Without the quartic isospin term, i.e. $q = 0$, the value of $J$ or $(1-q)E_{sym}(\rho_0)$ is $\approx 35$ MeV for AV8'+UIX. The role of quartic isospin term is to reduce the value of $J$ in the range $\approx 28$ to 30 MeV from 35 MeV, depending upon on other parameters such as $\rho_0$ and $u_v$ etc, through a suitable value of $q \neq 0$. The remaining strength, i.e. $qE_{sym}(\rho_0)$, is in the quartic term, contributes little because it multiplies $\delta^4$. We demonstrate this fact by calculating the percentage ratio of the quartic to quadratic isospin terms defined as

$$P = \frac{q}{1-q}\left(\frac{N-Z}{N+Z}\right)^2 \times 100 \qquad (16)$$

In Fig 5, we plot $P$ as a function of $A$ for 2149 nuclei for $q = 0.16$, a value representative of AV8'+UIX. It is seen that for more than 90% nuclei $P \leq 1\%$. The largest value of $P$ for the remaining nuclei is $< 2\%$. Thus it may appear that the strength of the quartic term is large ≈ 20% of the quadratic term, but its net contribution to the symmetry energy, thus to the total energy, is smaller by two orders of magnitude. Further, it also follows that the contribution of $\delta^6$ term in symmetry energy is an additional two orders of magnitude smaller than the values depicted in Fig 5. It is vanishingly small; this is the reason that we could not get any significant reduction in $\sigma(E)$ with the $\delta^6$ term.

Presence of a quartic term in symmetry energy at high densities has been proposed earlier [62] which strongly modifies critical density for the direct Urca process in connection with the cooling of neutron stars. We have demonstrated that its presence in the low density region is attributable to the use of EOS of neutron matter obtained through *ab initio* calculations with realistic interactions, in particular because of contribution from three-nucleon interaction near the equilibrium density. Some of the contribution also comes through clustering in the low density region since the clusters will have densities, at their centers, close to saturation density.

**B. Nuclear Matter Properties and Symmetry Energy:** In Table VIII, we give the relevant nuclear matter derived quantities. They are derived in the sense that they are the outcome of calculations presented in earlier tables. Column 2 refers to the tables and line number of earlier tables. For example, VI, 4 means that quantities in this row correspond to calculations and parameters of Table VI, line 4. We also give values of the constants *A*, *B*, *C* and *D* which appear in (6b). In the right panel of Fig 2, we plot the EOS for SNM for calculations given in Table VIII except for the results for VI, 4 for *K* = 260 MeV. The color code and the legends are given in the table caption. It is seen from the figure and Table VIII that for a change of $\rho_x$ by 0.02 fm$^{-3}$, the change in the location of maximum, $E_{max}$, in the EOS of SNM is only 0.005 fm$^{-3}$. Thus it is fixed around $\rho \approx 0.025$ fm$^{-3}$. We also show the EOS of Akmal *et al*. (APR) [26]. The triangles are the results with AV18 +UIX with relativistic boost corrections. The dotted line is obtained by invoking heuristic corrections to account for the known empirical values of $\rho_0$, $K$ and $u_v$. Below the spinodal density, around 0.05 fm$^{-3}$, all the many body calculations of SNM become unreliable; we do not have adequate techniques to address this problem reliably except perhaps the QS approach sheds some light [4]. At the moment, the present phenomenological approach seems a good alternative.

In Fig. 6, the results for the symmetry energies are given. The color code and legends for the various curves are same as those in Fig. 2 (right panel), except of the dotted curve. The dotted curve in Fig. 6 depicts the results of QS approach [2, 63] at T = 1 MeV which seems close to experiment. The experimental extraction of the symmetry energy was obtained in Ref. [2, 3], in the low density region in the pioneering experiment on heavy ion collisions of $^{64}Zn$ on $^{92}Mo$ and $^{197}Au$ at 35 MeV/$A$. The down blue triangles are the data from [2] obtained after correcting it for energy recalibration and reevaluation for particle yields in different velocity bins. They are therefore slightly different from [3]. We have shown an error bar of ±15% as reported in [3]. Symmetry energy is not a directly measurable quantity. It is extracted indirectly from other observables that depend on the symmetry energy, thus some model dependence or dependence on theoretical interpretations is inevitable. Significantly, the medium effects on the clusters play an important role [64, 65]. The up red triangles, are the data corrected for the medium effects in a self consistence way. The whole bunch of data points (down blue triangles) shifts to considerably higher densities (up red triangles) and there is an upward trend for the symmetry energies for lower densities; the down blue triangles have an opposite downward trend. The slopes of our calculated curves are negative at low densities as a result of our ansatz (6) and the EOS of neutron matter. This conforms to the data; the up red triangles which have been corrected for medium effects. Clearly, our calculations distinguish between the two sets of data, the up red and down blue triangles. Our symmetry energy shows a distinct minimum at $\rho_{min} \approx 0.02$ fm$^{-3}$. In QS approach, this minimum is not seen, because heavier clusters are not included. Thus, it is important that this region of density should be explored experimentally. The minimum in the symmetry energy occurs because it is a difference between EOS of pure neutron matter, which increases monotonically with density, and the EOS of SNM which has a maximum. Our symmetry energy curve for $\rho_x = 0.05$ fm$^{-3}$ seems to explain the experiment better, but we put more reliance on curves for $\rho_x = 0.06$ and $0.07$ fm$^{-3}$ because of the following two reasons. Firstly, theoretical curves are for T=0 MeV, whereas experimental points are at finite temperatures (average temperature is around 4.5 MeV). The symmetry energy is known to have some temperature dependence [2-4], becoming higher at lower temperatures. Secondly, the experimental extractions have been carried out by assuming no clusters beyond $A = 4$ which may not be a safe assumption. Presence of a heavier cluster will push the symmetry energy up. In Ref. [2], comparison of the data has also been made for the QS estimates at T = 4 and 8 MeV. One would expect that T = 4 MeV curve should be closer to data but this is not the case (see Fig.1 of Ref. 2). These curves show that symmetry energy decreases with density whereas experimentally, the up red triangles, show an upward trend. The temperature varies between 3.3 to 7.5 MeV for various densities. The open circles are calculations of the QS approach at the specific temperatures and densities as that of experiment. The QS points are taken from column 8 of Table I, Ref. [2].

The right panel of Fig 6 gives the experimental symmetry energy data as obtained from studies in heavy ion multi-fragmentation reactions at relatively higher densities. The red and green circles are from [66], the blue circle from [67] and the pink square is from [68] at the saturation density. It is seen that these data are satisfactorily explained in the present framework. These data have also been considered by Liu *at al* [69] with in the framework of LDM. A satisfactory fit in the density range $0.42\rho_0 \leq \rho \leq 0.64\rho_0$ was obtained. Other methods [70, 71] too give a reasonable account of these data. But none of these methods can account for the symmetry energy data of Natowitz *et al.* [2].

In Fig. 7, we plot the energy per nucleon of asymmetric nuclear matter, $E(\rho,\delta)$, for a number of values of the asymmetry parameter $\delta$. We use Eq. (4) (or equivalently Eq. (5)). For $E(\rho, \delta = 0)$ we use the values represented by the dashed (green) line in the right panel of Fig. 2 corresponding to EOS parameters of SNM for AV8' $+ V_{2\pi}^{pw} + V_{\mu=300}^{R}$ from Table VI. The vertical short lines in the figure indicate the location of equilibrium densities (drawn to aid the eyes), at which the energies show a minimum, for various values of the asymmetry $\delta$. The curve for $\delta = 1$ evidently represents the pure neutron matter EOS of the left panel of Fig. 2 corresponding to AV8' $+ V_{2\pi}^{pw} + V_{\mu=300}^{R}$ interaction. It is seen that the minimum in the energy disappears for $\delta$ somewhere between 0.8 and 0.9, i.e., they rise monotonically with density. We also bring to notice that all the curves meet the *y*-axis ($E(\rho, \delta)$-axis) at non-zero negative values for $\delta < 1$. Such a situation is not encountered in earlier nuclear matter calculations where $E(\rho \to 0, \delta) \to 0$. Importantly these negative values of energies (at $E(\rho, \delta)$-axis) are always less than the corresponding values of the minimum energies at the equilibrium densities for the particular values of $\delta$. They represent the ground state of the asymmetric nuclear matter. This is a consequence of the identity (1) and has the following interpretation. The equilibrium densities represent a stable state of the system but not the ground state. They correspond to uniform distribution of neutrons and protons throughout the nuclear volume. On the other hand the asymmetric nuclear matter in the limit of zero density can be considered as a blob of symmetric nuclear matter with saturation density $\rho_0$ consisting of Z protons and an equal number of neutrons surrounded by the remaining *N–Z* neutrons. A fraction of these remaining neutrons will be bound to the "periphery" of the SNM blob and rest will fly off with zero binding energies rendering the volume of the total system "infinitely more infinite" than the volume of the SNM blob which in itself is infinite. Such an asymmetric nuclear matter will always have a negative energy per nucleon as long as $\delta <$ 1. This negative energy corresponds to the ground state of the system. For example for $\delta = 0.8$, the energy per nucleon at the equilibrium density is positive, $\approx 1.7$ MeV, while in the limit of zero density it is negative, $\approx - 6$ MeV. We can calculate the energy per nucleon of the extra $N - Z$ neutrons which are outside the SNM blob. From (5), in the limit of zero density, we have

$$E(\rho \to 0, \delta) = E(\rho \to 0, \delta = 0) + \left(E(\rho \to 0, \delta = 1) - E(\rho \to 0, \delta = 0)\right)\left((1-q)\delta^2 + q\delta^4\right) \quad (17)$$

Recalling that $E(\rho \to 0, \delta = 0) \to -u_v$ and $E(\rho \to 0, \delta = 1) \to 0$ the above expression becomes

$$E(\rho \to 0, \delta) = -u_v + u_v \left((1-q)\delta^2 + q\delta^4\right) \quad (18)$$

Expression (18) is the energy per nucleon of the entire asymmetric nuclear matter in the limit of zero density. These are the energy values for the curves in Fig. 7 when they meet the $E(\rho, \delta)$-axis. The SNM blob consists of $Z$ protons and an equal number of $Z$ neutrons. Its energy per nucleon will be $-u_v 2Z/A$, where $A = N+Z$. We can express $Z$ in terms of $\delta$ as $Z = (1-\delta)A/2$. Substituting the value of $Z$, the energy per nucleon of the SNM blob becomes

$$E(\delta, \text{SNM blob}) = -u_v(1-\delta) \quad (19)$$

If we subtract (19) from (18), we shall obtain the energy per nucleon of the extra $N - Z$ neutrons:

$$E(\delta, \text{extra}(N-Z)\text{ neutrons}) = u_v(-\delta + (1-q)\delta^2 + q\delta^4) \quad (20)$$

Notice that (20) gives the correct limit, namely 0, when $\delta = 0$ or 1. In the former case, i.e., $\delta = 0$, there are no extra neutrons and in the latter case we have only neutrons at zero pressure and density. In Fig. 8 we plot (18) and (20) as a function of $\delta$. The lower curve, Eq. (18), gives the intersections of the curves in Fig. 7 when they meet at the $E(\rho, \delta)$-axis. It is seen that the excess $N - Z$ neutrons, represented by the uppermost curve, have considerable binding energies per nucleon with a maximum (or minimum in energy) around $\delta \approx 0.5$. The curve labeled as $E_{min}$ give the values of the minimum energies at the equilibrium densities as a function of $\delta$. In those approaches where $E(\rho \to 0) \to 0$, both the curves, the uppermost and the lowest, of Fig. 8 will be identically zero for all values of $\delta$, not a physically correct picture. It is thus not pure accident that we get significantly lower values of $rms$ deviations of the one neutron separation energies, $\sigma(S1)$, and particularly $\sigma(S_{nr})$, Table VII (result lines 3 and 6), as compared to those of Möller *et al.* Ref.[55] and Goriely *et al.*, Refs [56, 59]. In our theory, the tail region of neutron rich nuclei is better described than those in approaches where $E(\rho \to 0) \to 0$. No doubt, this is important for describing nuclei near the drip line.

A quantity of interest is the neutron skin thickness [72], defined as the difference between the $rms$ radii of neutrons and protons

$$\delta R = \sqrt{\langle r_n^2 \rangle} - \sqrt{\langle r_p^2 \rangle}. \tag{21}$$

In Skyrme-HF theories $\delta R$ is sensitive to the slope of the symmetry energy, $L$, at the saturation density, defined as

$$L = 3\rho_0 \left.\frac{\partial E_{sym}}{\partial \rho}\right|_{\rho_0}. \tag{22}$$

We expect the clustering to affect $\delta R$ significantly as it is a direct surface phenomenon. In Table IX, we give results of our calculations for clustering and no-clustering. Results for no-clustering are in reasonable agreement with Skyrme-HF [73, 74] and RMF [75] calculations, and experimental deductions [76, 79]. But there is clear discrepancy between the clustering and the other results including experiment. We find much lower values for $\delta R$. Does this imply that experimental deductions are implemented assuming no-clustering? They are indeed model dependent. Recently completed parity-violating electron scattering experiment [22] at the Jefferson Laboratory will greatly help to clarify this inconsistency. Our $L$ value for both the cases of clustering and no-clustering is same; $L \approx 68$ MeV well within the range of values extracted from isospin diffusion data.

Clearly, the structure of $E(\rho)$ for low densities is quite intricate which we have modeled through (6). The constant terms $A$, $B$, $C$ and $D$ in (6b-d) can be expressed in terms of $K$, $M$, $\rho_0$ and $\rho_x$ by equalizing $E(\rho)$ and its first three derivatives at $\rho_x$. This ensures that $K$ and its first derivative are continuous for $\rho \leq \rho_x$. This may not be so, for instance the compression modulus $K$ may not be even continuous at some low density. But this can only be decided through an accurate many-body calculation employing, for example, the AFDMC technique for SNM at low densities, or advancement in QS and other approaches [4] may shed some light. We have to wait till such calculations become feasible.

**C. Further Discussion:** A few comments on our approximate semi-classical energy density functional $\mathcal{E}[\rho_n, \rho_p]$, expression (7), are in order. The Hohenberg and Kohn theorem [80] holds for an exact density functional. Our approximate ETF functional will lead to a slight over binding problem if shell effects are not calculated in a self consistent manner particularly by the "expectation value method" as elaborated by Brack *et al.* [43], notwithstanding the fact that we have used a phenomenological pairing and Wigner terms. We believe this slight over binding, a few parts in $10^4$ for heavy nuclei [43], is easily compensated by the slight change in the parameters of the theory possibly indicated by the overall good fits to the static properties of nuclei. Further, a connection exists between the ETF energy functional and macroscopic phenomenological models such as liquid drop model and droplet model [43]. A quantitative connection has been established between the parameters of Skyrme forces which appear in the density functional with the parameters of the macroscopic models. Thus ETF energy functional embodies the macroscopic models; it is more general and powerful. Some of the macroscopic models with microscopic corrections [48, 50, 54, 55, 60, and 81] also suffer from the inconsistency problem mentioned earlier. Exceptions being the Duflo-Zuker [82] microscopic mass formula, recently developed fully microscopic density functional theories [11-13, 16], the latter also describe the spectroscopic properties of nuclei, and the microscopic-macroscopic theories, for example of Refs. [43, 44, 56]. Our σ($E$) values compete well with most of the other approaches. We have invoked ETF density functional to demonstrate that the low density symmetry energy data of Natowitz *et al.* [2] is consistent with the static properties of nuclei. We achieve this goal successfully. At the same token we have shown light on the properties of nuclear matter.

Next, we discuss the use of $E(\rho, \delta)$, (4), with $E(\rho)$ given by (6), in the density functional (7), an important ingredient in DFT [11-13, 83]. One often introduces a phenomenological term to control the surface properties [12, 83]. In our case, this term is the gradient $a_\rho (\nabla \rho)^2$ motivated by the Skyrme density functional. This does not affect the nuclear matter $E(\rho, \delta)$. Our $E(\rho)$ is phenomenological containing the empirically reasonable values of $K$, $u_v$ and $\rho_0$. We do not use any specific interaction, which in reality are very complicated, simply because to calculate $E(\rho)$ of SNM from any many body technique is a formidable task. This is even true for a simple Skyrme interaction; it will also give rise to clustering at low densities, thus generate strong correlations. One generally treats the Skyrme interaction as a kind of a phenomenological *G* matrix which includes the short range correlations. But the fact is that a perturbation calculation for second order would diverge because of zero range. We believe its justification lies mainly in its simplicity and ease of use (in the lowest order), and through the fact that it correlates large body of data. We could have also worked with a Skyrme density functional, within ETF formalism, by adding a phenomenological piece which guarantees that $E(\rho)$ tends to $-u_v$ as the density goes to zero in compliance with the identity (1). But, as demonstrated, we would have been confined to old FP [57] EOS of neutron matter as in the case of Refs. [56, 59] since Skyrme density functionals do not contain a quartic isospin term necessary for obtaining a good fit when use is made of the neutron matter EOS of Refs. [7, 8]. With the present density functional we thus stay close to the modern EOS of neutron matter which has been calculated reliably [7, 8] with realistic interactions [8, 14, 15].

Some comments on the parameters of the theory are required here. We discuss them one by one:

(a) Our values of *M*, the anharmonicity parameters, are negative and large in absolute term. We may point out that this parameter is imprecisely determined. We place an uncertainty of around 20%. With clustering they lie in the range of -9 to -17 MeV (Tables II, III, IV, and VI). For the APR EOS [26], dotted curve in the right panel of Fig.2, $M = -3.51$ MeV. But this curve has been obtained through a heuristic, though ad-hoc, corrections to the actual calculations represented by triangles, to account for the empirical properties of the SNM. APR assume $\rho_0 \approx 0.16$ fm$^{-3}$, $u_v \approx 16$ MeV and $K \approx 260$ MeV as empirical values of these parameters presumably not giving consideration to *M* values in the low density region as their emphasis is primarily in the EOS of NM at higher densities. With other nuclear interactions of Argonne-Urbana collaboration the |*M*| values

are also small (for details see section IIB of Ref. [17]). The effective interactions, Skyrme as well as finite range, follow the same trend. Remarkably, when we take away the clustering, using Eq. (6d), the $M$ values range between $\approx$ –5 MeV (Table V) and $\approx$ 4 MeV (result line 2 of Table VI). Considering the uncertainties in our determination of $M$, these values are then in agreement with the EOS calculations with realistic and effective interactions which do not have clustering in them. May be a more flexible (6b) solve this dichotomy. In the absence of any clear answer we do not dwell any further on this problem.

(b) The other parameter which occurs in (6) is $\rho_x$. It is obvious why his parameter is required. This is to connect the two segments of the EOS (6) to facilitate clustering. Its value is more or less fixed by the symmetry energy data of Natowitz *et al*. There is no counterpart to this parameter in the microscopically calculated EOS.

(c) The pairing term, (10), has a $1/A^{(1/3)}$ dependence and some dependence on isospin. This term has emerged through an "anatomy" [49] of the Duflo-Zuker [82] mass formula. It has been successfully employed by Wang *et al*. [50] and Liu *et al*. [60] in their excellent fit of nuclear binding energies. Our values of the parameter $a_{pair}$ are in the range $\approx$ –5.5 to –6.5 MeV in agreement with Wang *et al*. [50] and Liu *et al*. [60] which they respectively obtain as -5.5108 MeV and -6.2299 MeV. We may remark here that isospin dependence of the pairing energy is not important. It is the $1/A^{(1/3)}$ dependence which is significant.

(d) Parameters of the Wigner term, (11), are also on the expected lines. The two parameters $\lambda$ and $A_0$ have considerable spread since they are large and occur in the exponential with a minus sign. With the inclusion of clustering, $V_w$ varies between $\approx$ -3.0 to 0.9 MeV and $W_w$ between $\approx$ -0.4 to 0.26 MeV for the various fits. Goriely *et al*. [56] in their Skyrme-Hartree-Fock-Bogoliubov nuclear mass formulas find $V_w$ = -2 MeV and $W_w$

= 0.86 MeV. We may remark here that in a fully microscopic theory the Wigner term should not occur. In macroscopic models as well as in Refs. [51, 56 and 59], it is included as a phenomenological term for a precise fit with hardly any theoretical justification. Its weight in the binding energies of all the nuclei taken together is small; its parameters, the combination of four parameters in (11), are not precisely determined.

(e) The coefficient of the gradient term, $a_\rho$, varies between ≈ 37 to 40 MeV with clustering (Tables II to VI). It is difficult to compare its value with other approaches because of the various other gradient terms which occur in those and our approach as well as the difference in formulations. For example in [56], its value is ≈ 20 MeV, where as in [84] which is based upon ETF formulation its value is ≈ 32 MeV.

(f) Lastly, our values of $u_v$ and $\rho_0$ treated as free parameters are generally consistent with all other approaches and the saturation properties of nuclei. We did not treat $K$ as a free parameter. Its value was fixed at $K$ = 230 MeV in compliance with the findings of Ref. [23]. Except in one case where we did the calculations for $K$ = 260 MeV to demonstrate that this parameter is not sensitive to our results and conclusions. Few authors do prefer a larger value of $K$.

The purpose of the foregoing discussion was to emphasize that each term and the parameter of our theory has a physical origin except perhaps the Wigner component which eludes any convincing theoretical justification other than it gives a better fit. We have argued that the lack of self consistency of the type mentioned at a few places in this work is of not serious consequence. They are of the same type as they occur in some of the microscopic-macroscopic models.

## V CONCLUSIONS AND FUTURE OUTLOOK

In conclusion, we have presented a unified theory of nuclei which is consistent with the static properties of nuclei and clustering at the nuclear surface and incorporates the large values of the symmetry energies at low densities. In addition, it includes the EOS of pure neutron matter obtained through *ab initio* calculations with realistic interactions. Main conclusions are as follows:

(a) The slope of the symmetry energy is negative at low densities and positive at and near the saturation density which leads to a minimum in the symmetry energy.

(b) Establishes that quartic term in isospin plays an important role; it originates mainly from the use of realistic EOS of neutron matter, particularly contribution arising from three-neutron interaction, and somewhat from clustering.

(c) We have given estimates of the spinodal density, the density at which the symmetry energy has a minimum and the density at which the EOS of SNM has a maximum. Cluster formation in the SNM begins somewhere between the spinodal density $\rho_{sp} \approx$ 0.05 fm$^{-3}$ and $\rho_{max} \approx$ 0.025 fm$^{-3}$ and the symmetry energy has a minimum at $\rho \approx$ 0.02 fm$^{-3}$ below which larger clusters dominate.

(d) Clustering at low densities significantly reduces the neutron skin-thickness in nuclei. The conventional Skyrme density functionals do not incorporate clustering at low densities; thus their estimates of neutron skin-thickness can be considerably off. Also, since they do not contain quartic isospin term, it would be difficult to explain quantitatively the nuclear binding energies in a fully microscopic density functional theories and *at the same time conform to the realistic EOS of neutron matter, particularly in situations where they are steep, for example with* UIX.

(e) Lastly, the most important conclusion of the present study is that the EOS of SNM does not follow the trend $E(\rho) \to 0$ as $\rho \to 0$. Instead it follows the behavior as depicted in the right panel of Fig 2 which incorporates the identity (1). This implies that $E_{sym}(\rho \to 0) = u_v$. Also, this leads to an interesting and physically sound interpretation of the EOS of asymmetric nuclear matter. This in turn implies, as explained briefly in section IV that our formulation and considerations will lead to an accurate account of neutron rich nuclei.

Clearly, it is imperative that we should generalize the calculations which include deformation, shell and other microscopic corrections in a self consistent manner with in the present framework. The next step would be to develop a fully microscopic theory as envisioned in the SciDAC Review [16] incorporating important features of the present study.

## VI ACKNOWLEDGEMENT

QNU acknowledges useful discussion with S. C. Pieper and R. B. Wiringa regarding the various aspects of the EOS of SNM. He is grateful to S. C. Pieper for suggesting the inclusion of EOS of $^4He$ liquid in section II. Authors gratefully acknowledge correspondence with S. Shlomo, G. Röpke, K. E. Schmidt and A. W. Steiner. They are thankful to the authorities of University Malaysia Perlis for providing the computing facility at the Institute of Engineering Mathematics where this work was performed. We gratefully acknowledge the University Malaysia Perlis research incentive grant no. 9007-00025.

| Clustering | $\sigma(E)$ MeV | $\sigma(R)$ fm | $a_\rho$ MeV fm$^5$ | $a_{np}$ MeV fm$^5$ | $-a_{pair}$ MeV | $-M$ MeV | $\rho_0$ fm$^{-3}$ | $u_v$ MeV | $-V_w$ MeV | $\lambda$ | $-W_w$ MeV | $A_0$ |
|---|---|---|---|---|---|---|---|---|---|---|---|---|
| Yes (6b) | 1.489 | 0.022 | 40.328 | 0.0 | 7.722 | 18.368 | 0.1539 | 16.000 | 18.71 | 44 | 1.76 | 480 |
|  | 1.091 | 0.021 | 38.395 | 27.347 | 7.234 | 17.536 | 0.1542 | 15.937 | 13.27 | 56 | 1.34 | 653 |
| Yes (6c) | 1.578 | 0.026 | 38.994 | 0.0 | 7.404 | 21.353 | 0.1545 | 16.042 | 20.37 | 44 | 1.89 | 453 |
|  | 1.335 | 0.021 | 39.232 | 18.991 | 7.628 | 19.932 | 0.1538 | 15.994 | 16.92 | 51 | 1.60 | 482 |

Table I : Results with $q = 0$ and $\rho_x = 0.06$ fm$^{-3}$. Neutron matter EOS is with AV8'+UIX. Clustering is included using (6c) and (6b) in the EOS of SNM. Notice the significant reduction in $\sigma_E$ when $a_{np}$ is varied compared to $a_{np}= 0$.

| Clustering | $\sigma(E)$ MeV | $\sigma(R)$ fm | $q$ | $(1-q)E_{sym}(\rho_0)$ MeV | $qE_{sym}(\rho_0)$ MeV | $a_\rho$ MeV fm$^5$ | $-a_{pair}$ MeV | $-M$ MeV | $\rho_0$ fm$^{-3}$ | $u_v$ MeV | $-V_w$ MeV | $\lambda$ | $-W_w$ MeV | $A_0$ |
|---|---|---|---|---|---|---|---|---|---|---|---|---|---|---|
| Yes(6b) | 0.589 | 0.020 | 0.168 | 29.76 | 6.01 | 38.709 | 5.912 | 13.410 | 0.1586 | 15.970 | 2.98 | 200 | 0.343 | 68 |
| Yes(6c) | 0.584 | 0.020 | 0.170 | 29.70 | 6.08 | 38.544 | 5.850 | 14.386 | 0.1586 | 15.974 | 2.77 | 208 | 0.297 | 67 |

Table II: Results with $a_{np}= 0$ and $\rho_x = 0.06$ fm$^{-3}$ and the quartic isospin term. There is a dramatic reduction in $\sigma_E$ compared to table I. Neutron matter EOS is with AV8'+UIX.

| Neutron Matter | Clustering | $\sigma(E)$ MeV | $\sigma(R)$ fm | $q$ | $(1-q)E_{sym}(\rho_0)$ MeV | $qE_{sym}(\rho_0)$ MeV | $\rho_x$ fm$^{-3}$ | $a_\rho$ MeV fm$^5$ | $-a_{pair}$ MeV | $-M$ MeV | $\rho_0$ fm$^{-3}$ | $u_v$ MeV | $-V_w$ MeV | $\lambda$ | $-W_w$ MeV | $A_0$ |
|---|---|---|---|---|---|---|---|---|---|---|---|---|---|---|---|---|
| AV8'+UIX | Yes(6c) | 0.600 | 0.021 | 0.163 | 29.79 | 5.80 | 0.04 | 37.742 | 5.891 | 11.111 | 0.1576 | 15.919 | 2.96 | 213 | 0.387 | 61 |
|  | Yes(6b) | 0.595 | 0.021 | 0.165 | 29.74 | 5.88 | 0.05 | 38.182 | 5.949 | 12.099 | 0.1577 | 15.938 | 3.04 | 188 | 0.362 | 68 |
|  | Yes(6c) | 0.610 | 0.021 | 0.166 | 29.69 | 5.91 | 0.05 | 37.404 | 5.652 | 12.671 | 0.1577 | 15.920 | 2.51 | 326 | 0.341 | 51 |
|  | Yes(6b) | 0.578 | 0.021 | 0.168 | 29.52 | 5.96 | 0.07 | 37.201 | 5.856 | 16.343 | 0.1566 | 15.942 | 2.78 | 224 | 0.277 | 70 |
|  | Yes(6c) | 0.583 | 0.021 | 0.172 | 29.50 | 6.13 | 0.07 | 37.332 | 5.740 | 17.272 | 0.1576 | 15.960 | 2.44 | 269 | 0.220 | 63 |
| AV8' + $V_{2\pi}^{PW}$ + $V_{\mu=300}^R$ | Yes(6b) | 0.575 | 0.020 | 0.093 | 28.88 | 2.96 | 0.06 | 38.463 | 5.782 | 12.957 | 0.1589 | 15.946 | 2.58 | 336 | 0.328 | 64 |

Table III: Results with different values of $\rho_x$.

| Neutron Matter | Clustering | $\sigma(E)$ MeV | $\sigma(R)$ fm | $q$ | $(1-q)E_{sym}(\rho_0)$ MeV | $qE_{sym}(\rho_0)$ MeV | $a_\rho$ MeV fm$^5$ | $-a_{pair}$ MeV | $-M$ MeV | $\rho_0$ fm$^{-3}$ | $u_v$ MeV | $-V_w$ MeV | $\lambda$ | $-W_w$ MeV | $A_0$ |
|---|---|---|---|---|---|---|---|---|---|---|---|---|---|---|---|
| AV8' | Yes(6b) | 0.570 | 0.020 | 0.082 | 28.27 | 2.53 | 38.178 | 5.776 | 13.093 | 0.1585 | 15.934 | 2.60 | 347 | 0.320 | 68 |
|  | Yes(6c) | 0.572 | 0.020 | 0.084 | 28.13 | 2.58 | 37.751 | 5.724 | 14.289 | 0.1584 | 15.940 | 2.45 | 365. | 0.283 | 67 |

Table IV: Results for $\rho_x = 0.06$ fm$^{-3}$ with AV8'.

| Neutron Matter | Clustering | $\sigma_E$ MeV | $\sigma_R$ fm | $q$ | $(1-q)E_{sym}(\rho_0)$ MeV | $qE_{sym}(\rho_0)$ MeV | $a_\rho$ MeV fm$^5$ | $a_{np}$ MeV fm$^5$ | $-a_{pair}$ MeV | $-M$ MeV | $\rho_0$ fm$^{-3}$ | $u_v$ MeV | $-V_w$ MeV | $\lambda$ | $-W_w$ MeV | $A_0$ |
|---|---|---|---|---|---|---|---|---|---|---|---|---|---|---|---|---|
| AV8'+UIX | No(6d) | 0.637 | 0.024 | 0.150 | 30.17 | 5.32 | 42.224 | 0.0 | 5.904 | 5.198 | 0.1571 | 15.887 | 7.41 | 21.3 | 0.322 | 68 |
| AV8' |  | 0.622 | 0.019 | 0.062 | 28.85 | 1.91 | 41.420 | 0.0 | 5.954 | 5.374 | 0.1581 | 15.887 | 7.80 | 13.8 | 0.231 | 76 |
|  |  | 0.757 | 0.018 | 0.0 | 30.73 | 0.00 | 41.161 | 75.719 | 5.865 | 5.260 | 0.1576 | 15.883 | 5.78 | 29.9 | -0.472 | 54 |

Table V: Results with no-clustering and for $\rho_x = 0.06$ fm$^{-3}$. Increase in $\sigma_E$, because of no-clustering situation, is 8.5% for AV8'+UIX and 7.8% for AV8' (computed from the entries of this Table and the Tables II, III and IV). For AV8' + $V_{2\pi}^{PW}$ + $V_{\mu=300}^R$, the increase is 6%.

| Neutron Matter | Clustering | $\sigma(E)$ MeV | $\sigma(R)$ fm | $\sigma(S1)$ MeV | $\sigma(S2)$ MeV | $\sigma(Q_\beta)$ MeV | $q$ | $(1-q)E_{sym}(\rho_0)$ MeV | $qE_{sym}(\rho_0)$ MeV | $a_\rho$ MeV fm$^5$ | $-a_{pair}$ MeV | $-M$ MeV | $\rho_0$ fm$^{-3}$ | $u_v$ MeV | $-V_w$ MeV | $\lambda$ | $-W_w$ MeV | $A_0$ |
|---|---|---|---|---|---|---|---|---|---|---|---|---|---|---|---|---|---|---|
| AV8'+UIX | Yes (6b) | 0.664 | 0.044 | 0.391 | 0.519 | 0.510 | 0.171 | 29.31 | 6.04 | 40.672 | 6.244 | 12.273 | 0.1561 | 15.888 | 2.63 | 75 | 0 | – |
| AV8'+UIX | No(6d) | 0.671 | 0.042 | 0.419 | 0.564 | 0.557 | 0.151 | 29.87 | 5.31 | 43.676 | 5.877 | -4.646 | 0.1551 | 15.840 | 8.23 | 22 | 0.171 | 91 |
| AV8'+$V_{2\pi}^{PW}$+$V_{\mu=300}^{R}$ | Yes(6b) | 0.650 | 0.043 | 0.381 | 0.508 | 0.506 | 0.097 | 28.36 | 3.05 | 38.444 | 6.134 | 12.420 | 0.1552 | 15.817 | 1.50 | 391 | -0.079 | 129 |
| AV8' | Yes(6b) | 0.637 | 0.043 | 0.380 | 0.505 | 0.504 | 0.086 | 27.90 | 2.62 | 38.523 | 6.227 | 12.483 | 0.1552 | 15.822 | 1.76 | 373 | -0.034 | 186 |
| No Wigner term | | 0.757 | 0.043 | 0.430 | 0.607 | 0.619 | 0.085 | 27.92 | 2.62 | 38.532 | 6.481 | 12.481 | 0.1552 | 15.822 | 0.0 | – | 0.0 | – |
| (*)AV8'+UIX | Yes (6b) | 0.665 | 0.044 | 0.405 | 0.529 | 0.545 | 0.195 | 28.34 | 6.87 | 39.614 | 5.480 | 9.501 | 0.1552 | 15.853 | 0.87 | 1341 | -0.258 | 429 |

Table VI: Results with 2149 nuclei. $\rho_x$=0.06 fm$^{-3}$. $\sigma_{S1}$, $\sigma_{S2}$, and $\sigma(Q_\beta)$ are for the measured 1988, 1937 and 1868 nuclei respectively. The result line corresponding to $\sigma_E = 0.757$ for AV8' is without the Wigner term. The row with asterisk (*) before AV8'+UIX is for results with $K$=260 MeV demonstrating a small dependence on $K$.

| rms deviations | Present | Möller et. al. Ref.[55] | Brussels-Montreal | | LDM | | LDM+WK Ref. [54] |
|---|---|---|---|---|---|---|---|
| | | | HFB-14 Ref.[59] | HFB-17 Ref.[56] | Ref. [50] | Ref. [60] | |
| $\sigma(E)$ MeV | 0.570 0.637 | 0.669 | 0.729 | 0.581 | 0.516 | 0.360 | 0.630 |
| $\sigma(R)$ fm | 0.020 0.043 | – | 0.031 | 0.030 | – | – | – |
| $\sigma(S1)$ MeV | 0.360 0.380 | 0.411 | – | 0.506 | 0.346 | 0.248 | – |
| $\sigma(S2)$ MeV | 0.492 0.505 | – | – | – | – | – | – |
| $\sigma(Q_\beta)$ MeV | – 0.504 | – | – | 0.583 | – | – | – |
| $\sigma(S_{nr})$ MeV | – 0.561 | 0.910 | 0.833 | 0.729 | – | – | – |
| No. of Nuclei | 367 2149 | 1654 | 2149 | 2149 | 2149 | 2149 | 367 |

Table VII: Root mean square deviations in various approaches compared with the present calculations. The *rms* deviation $\sigma(S_{nr})$ is for the subset of 186 neutron rich nuclei for which the one neutron separation energy is < 5 MeV. The last line gives the number of nuclei included in the fits. For further details, see text.

| Neutron Matter | Ref. Table & Line No. | $\rho_x$ fm | $\rho_{min}$ fm | $\rho_{max}$ fm | $\rho_{sp}$ fm | $E_{max}$ MeV | $A\times 10^{-3}$ MeV fm$^3$ | $-B\times 10^{-5}$ MeV fm$^6$ | $C\times 10^{-6}$ MeV fm$^9$ | $-D\times 10^{-7}$ MeV fm$^{12}$ |
|---|---|---|---|---|---|---|---|---|---|---|
| $AV8'+V_{2\pi}^{PW}+V_{\mu=300}^{R}$ | VI, 3 | 0.06 | 0.020 | 0.025 | 0.048 | -2.08 | 0.61281 | 0.31641 | 0.38446 | 0.16377 |
| AV8' | VI, 4 | 0.06 | 0.020 | 0.025 | 0.044 | -2.05 | 0.61582 | 0.31738 | 0.38561 | 0.16425 |
| AV8'+UIX | VI, 1 | 0.06 | 0.020 | 0.025 | 0.048 | -2.18 | 0.60807 | 0.31436 | 0.30183 | 0.16263 |
|  | VI, 6 | 0.06 | 0.020 | 0.025 | 0.044 | -2.52 | 0.56443 | 0.29855 | 0.36215 | 0.15415 |
|  | III, 2 | 0.05 | 0.018 | 0.022 | 0.041 | -0.96 | 0.89402 | 0.47174 | 0.67233 | 0.34173 |
|  | III, 4 | 0.07 | 0.022 | 0.027 | 0.053 | -1.63 | 0.55874 | 0.23698 | 0.28032 | 0.10295 |

Table VIII: Symmetric nuclear matter derived quantities. For details see text.

| Nucleus | $\delta R$ (fm), neutron skin thickness | | | | | |
|---|---|---|---|---|---|---|
| | Clustering | No-clustering | HFB-17 [56] | Sky-HF [73,74] | RMF [75] | Experiment |
| $^{208}Pb$ | 0.10 | 0.16 | 0.15 | 0.22±0.04 | 0.21 | 0.16±0.06 [77] |
| | | | | | | 0.18±0.035 [78] |
| $^{132}Sn$ | 0.15 | 0.23 | – | 0.29±0.04 | 0.27 | 0.24±0.04 [78] |
| $^{124}Sn$ | 0.10 | 0.16 | – | 0.22±0.04 | 0.19 | 0.185±0.017 [79] |
| $^{90}Zr$ | 0.04 | 0.07 | – | 0.09±0.04 | – | 0.07±0.04 [76] |
| $^{48}Ca$ | 0.10 | 0.16 | – | – | – | – |

Table IX: Results for neutron skin thickness for a number of nuclei. All entries are in fermis. Results for no-clustering correspond to the second row of Table VI. Results for clustering correspond to the averages of the first, third and the fourth rows of Table VI. Notice the appreciable reduction in $\delta R$ when clustering is included.

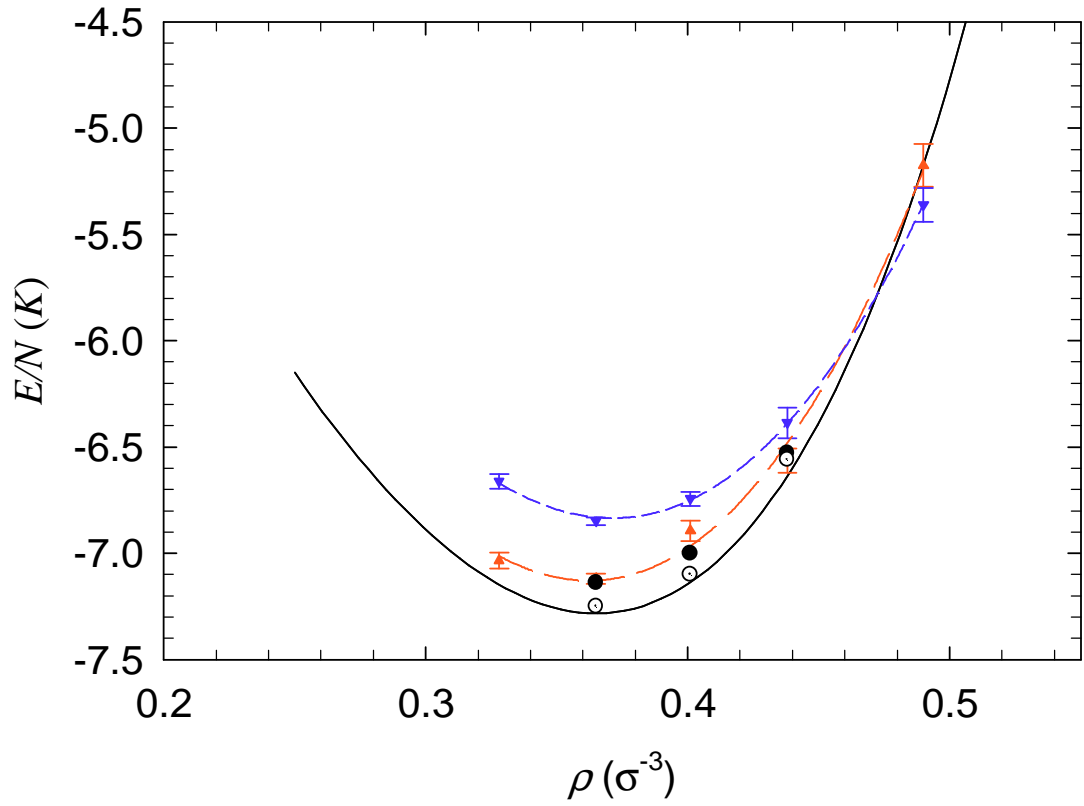

Fig. 1 (Color online): Equation of state of liquid $^4He$ as a function of density demonstrating its saturation character. The saturation density is 0.365 $\sigma^{-3}$ below which the energy rises with lowering of density. For details, see text.

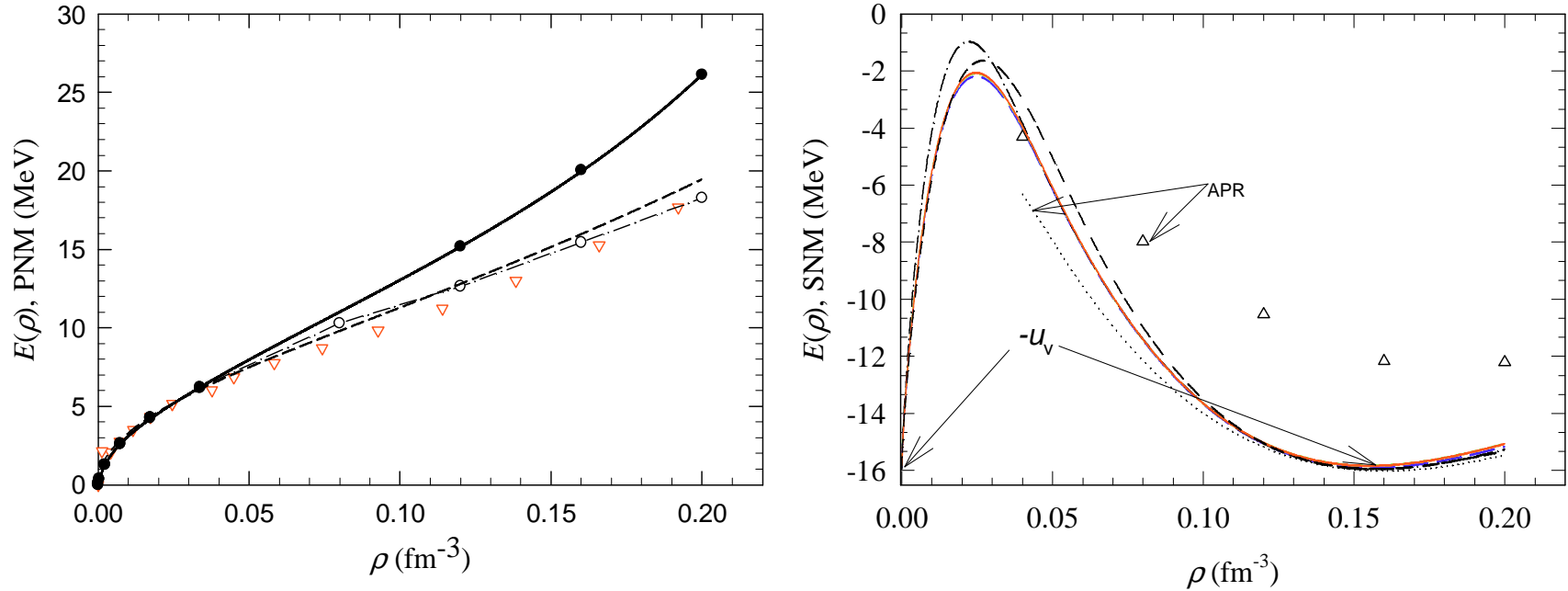

Fig.2 (Color online): Equation of State of neutron and SNM as function of density. **Left panel** (Neutron Matter): The filled circles are the results of AFDMC calculations with AV8'+UIX of Ref. [7]. The open circles are results with AV8' alone; dot-dashed line (to join the open circles) is drawn to aid the eye. The solid line is the fit obtained through Pade approximation as described in the text. The dashed line is the results for AV8' $+V_{2\pi}^{PW}+V_{\mu=300}^{R}$ as described in the text, Ref. [8]. The down triangles (FP) are from Ref. [57]. **Right panel** (Symmetric Nuclear Matter): The long dashed line (blue) represents the result for $AV8'+UIX$, the solid line (red) for $AV8'$, and the dashed line (green) for AV8' $+V_{2\pi}^{PW}+V_{\mu=300}^{R}$. All these curves are for $\rho_x = 0.06$ fm$^{-3}$ and almost merge with each other. The dashed-dot and short-short curves are for AV8'+UIX with $\rho_x = 0.05$ fm$^{-3}$ and $\rho_x = 0.07$ fm$^{-3}$ respectively. The triangles and the dotted line are the calculations of APR [26].

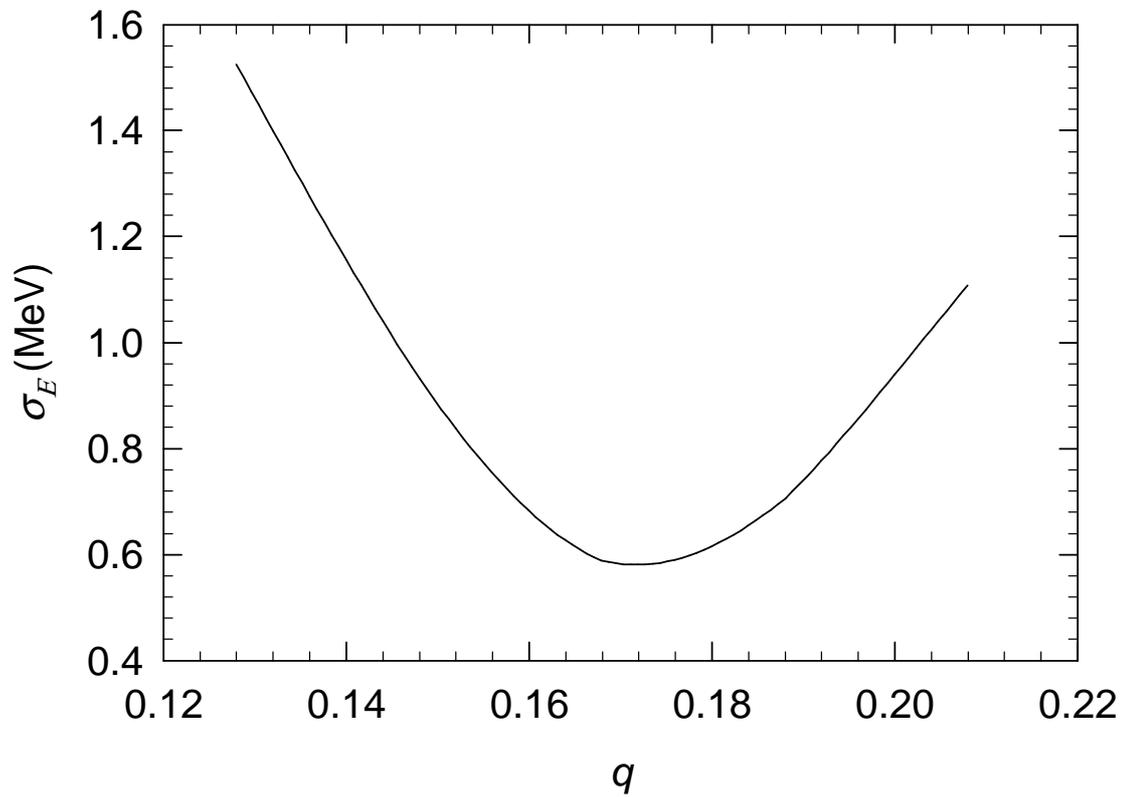

Fig 3: $\sigma(E)$ as a function of $q$. Neutron matter EOS is with AV8'+UIX and SNM is with (6b).

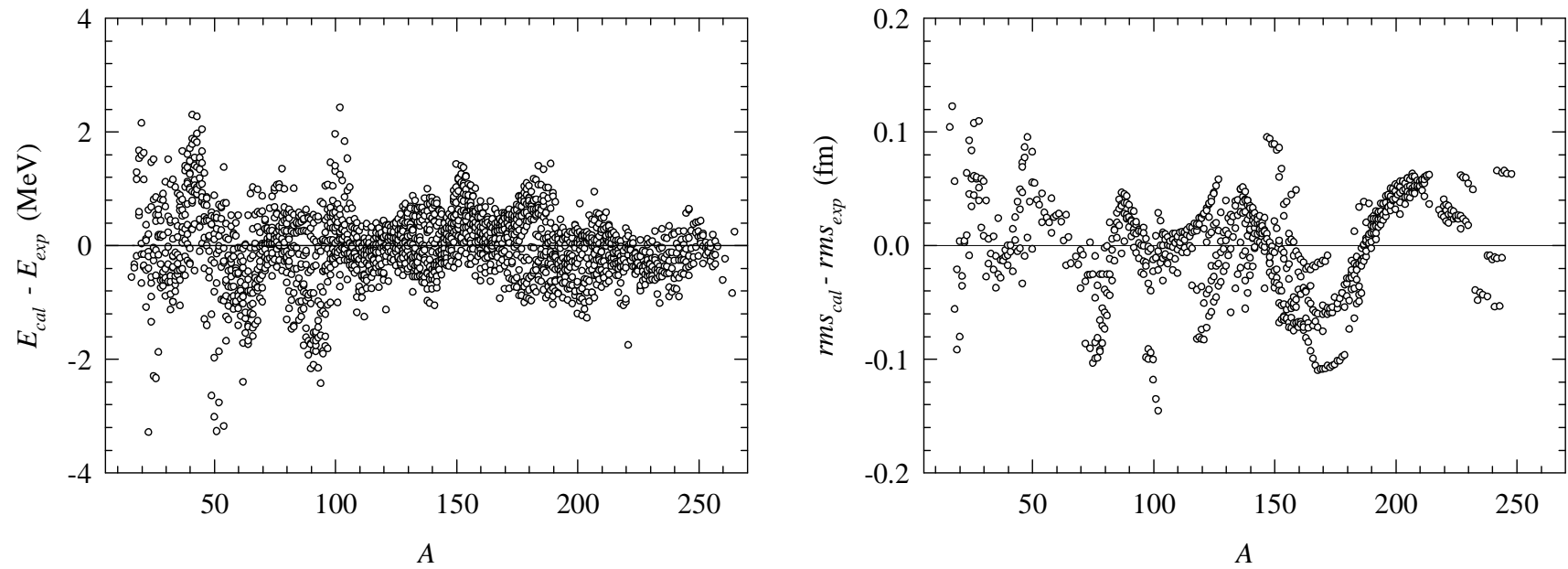

Fig. 4: Calculated and experimental differences for energies (left panel, 2149 nuclei) and proton point *rms* radii (right panel, 773 nuclei) as a function of $A$. The points are for AV8' from table VI.

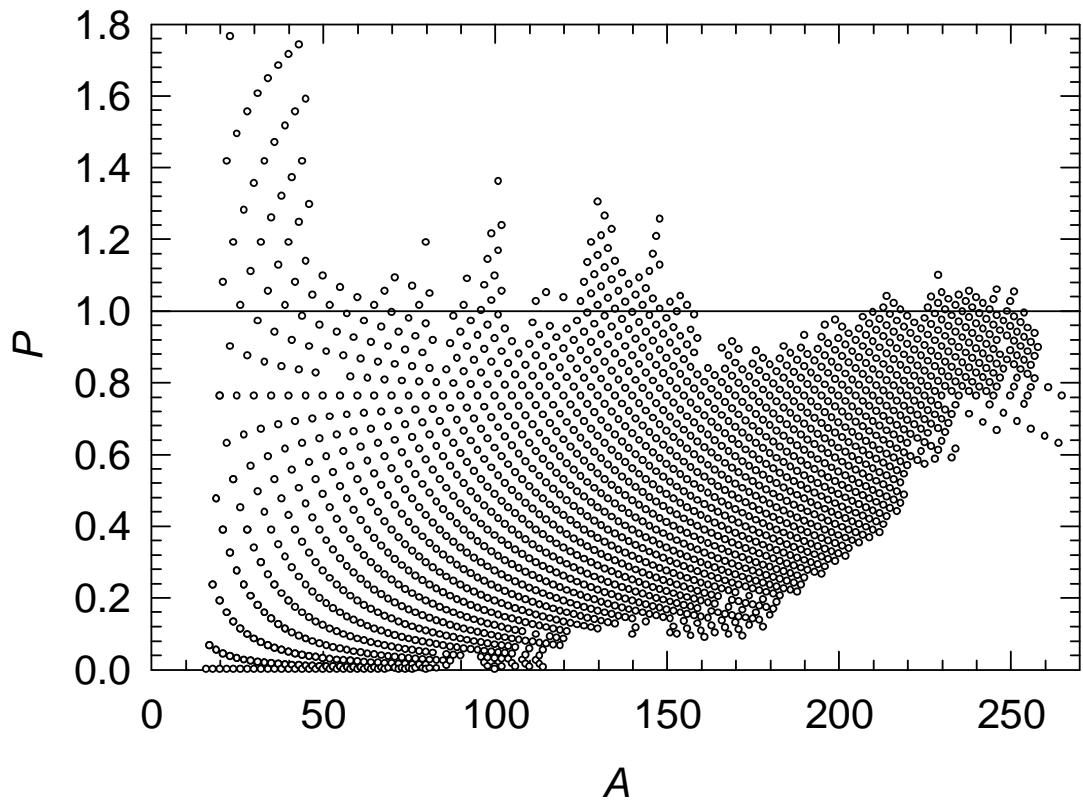

Fig 5: Percentage ratio of quartic to quadratic isospin terms of the symmetry energy plotted as a function of mass number $A$ for 2149 nuclei for $q = 0.16$.

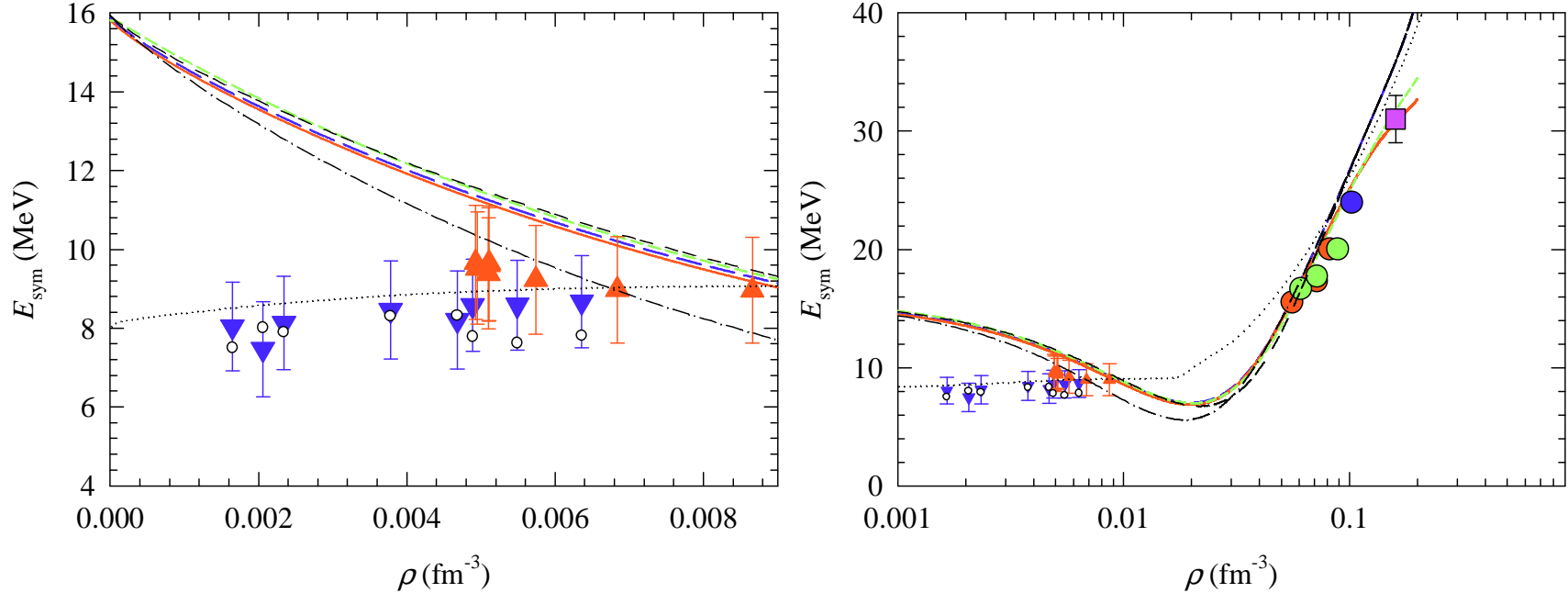

Fig. 6 (Color online): Symmetry energy as function of density. The long dashed line (blue) represents the result for $AV8'+UIX$, the solid line (red) for $AV8'$, and the dashed line (green) for $AV8'+V_{2\pi}^{PW}+V_{\mu=300}^{R}$. All these curves are for $\rho_x = 0.06$ fm$^{-3}$. The dashed-dot and the short-short curves are for $AV8'+UIX$ with $\rho_x = 0.05$ fm$^{-3}$ and $\rho_x = 0.07$ fm$^{-3}$, respectively. The up (red), with medium modifications, and down (blue) triangles are the data points from Ref. [2]. The dotted line depicts the results of QS approach [2, 63] at T = 1 MeV. The open circles are the results of the QS calculations at the specific densities and temperatures; taken from column 8 of Table I of Ref. [2]. See text for details for the data points represented by square and circles in the right panel.

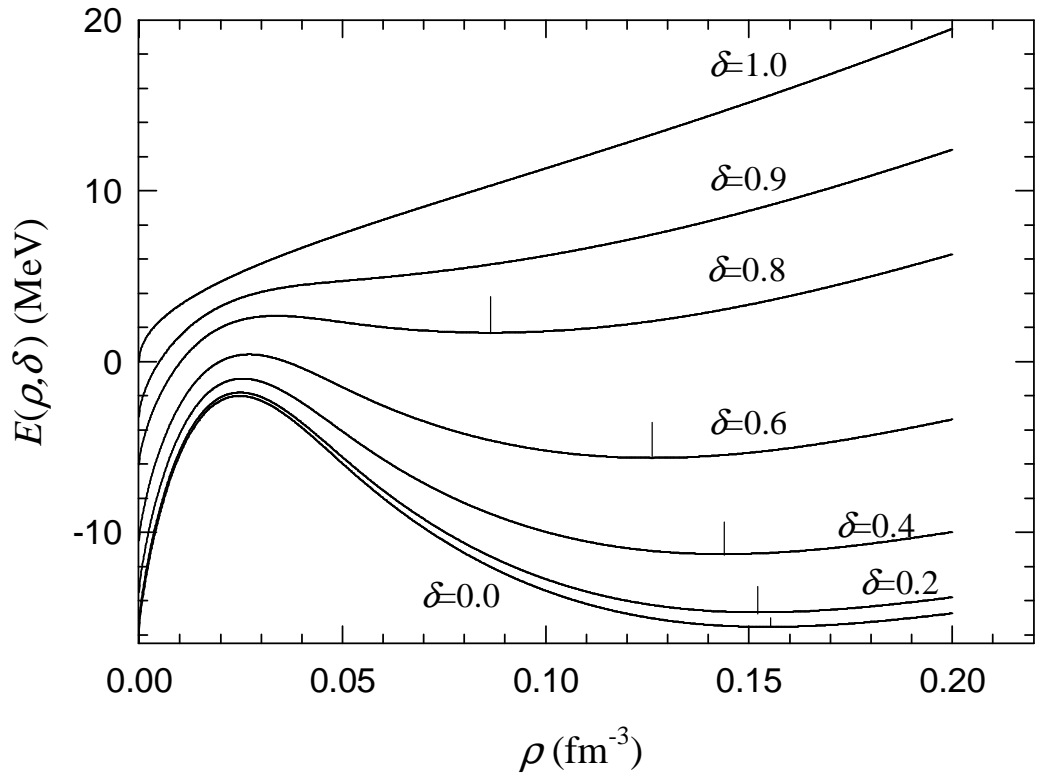

Fig. 7: Equation of state of asymmetric nuclear matter as a function of density for different asymmetry parameter $\delta$. The vertical short lines indicate the location of minima at the equilibrium density. For details see text.

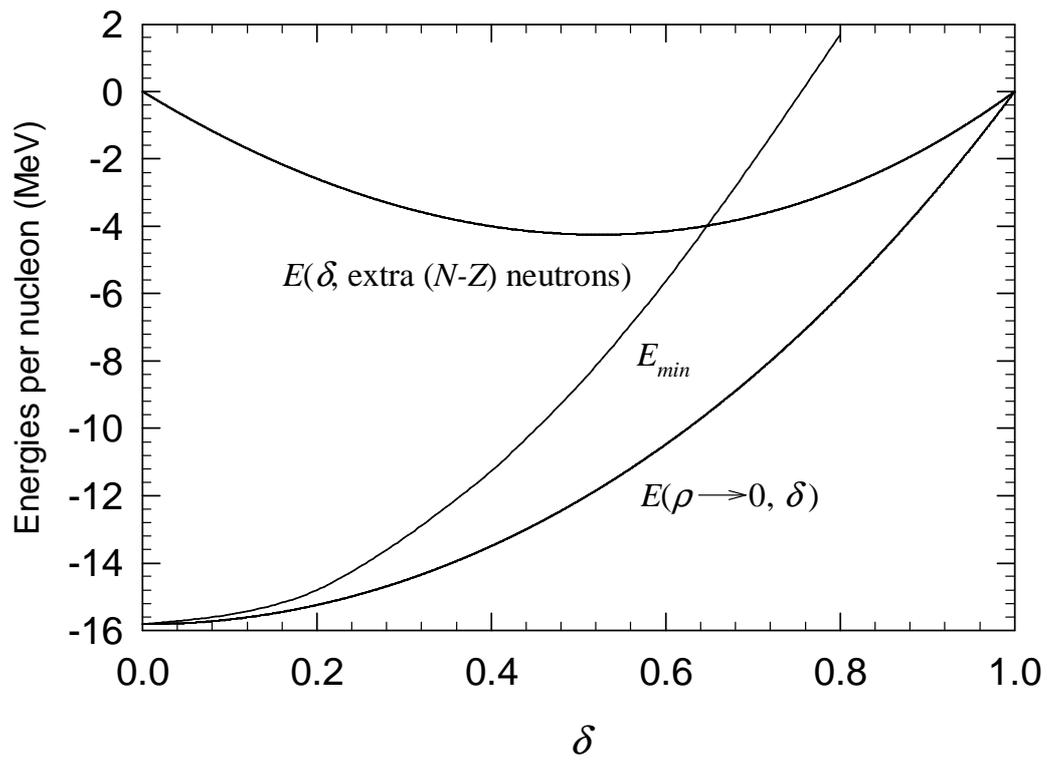

Fig 8: The curve labeled as $E(\rho \rightarrow 0, \delta)$ gives ground state energies of the asymmetric nuclear matter as function of $\delta$. The curve $E(\delta,$ extra $(N\text{-}Z)$ neutrons$)$ is the energy of excess neutrons and the curve labeled $E_{min}$ gives the minimum energies at the equilibrium densities of Fig. 7 as a function of $\delta$. All energies are per nucleon.